\RequirePackage{ifpdf}
\ifpdf 
\documentclass[pdftex]{sigma}
\else
\documentclass{sigma}
\fi

\usepackage{lscape}

\newcommand\C{{\mathbb C}}
\newcommand\N{{\mathbb N}}
\newcommand\R{{\mathbb R}}
\newcommand\D{{\mathbb D}}

\newcommand\cB{{\mathcal B}}
\newcommand\cC{{\mathcal C}}
\newcommand\cO{{\mathcal O}}
\newcommand\cT{{\mathcal T}}
\newcommand\ps{{\mathcal P}}
\newcommand\cR{{\mathcal R}}
\newcommand\cE{{\mathcal E}}

\newcommand\bm{{\mathbf m}}                   
\newcommand\Aut{\operatorname{Aut}}
\renewcommand\Re{\operatorname{Re}}

\newcommand\be{\begin{equation}}
\newcommand\ee{\end{equation}}

\numberwithin{equation}{section}

\numberwithin{theorem}{section}
\numberwithin{proposition}{section}
\numberwithin{lemma}{section}
\numberwithin{corollary}{section}
\numberwithin{definition}{section}
\numberwithin{example}{section}
\numberwithin{remark}{section}

\newcommand\ow{\overline w}
\newcommand\oz{\overline z}
\newcommand\dbar{\overline\partial}

\let\epsilon\varepsilon
\let\kappa\varkappa

\newcommand\wt{\widetilde}
\newcommand\unu{U^{(\nu)}}
\renewcommand\AA{\mathcal A}
\newcommand\jedna{\mathbf 1}
\newcommand\TT{\mathcal T}
\newcommand\WW{\mathcal W}
\newcommand\knu{K^{(\nu)}}
\newcommand\bn{{\mathbf n}}
\newcommand\AAC{\AA^\C}
\newcommand\cP{\mathcal P}

\newcommand\qH{\mathbb H}
\newcommand\Ka{\mathbb K}
\newcommand\bO{{\mathbb O}}

\begin{document}

\allowdisplaybreaks

\renewcommand{\thefootnote}{$\star$}

\renewcommand{\PaperNumber}{021}

\FirstPageHeading

\ShortArticleName{Toeplitz Quantization and Asymptotic
 Expansions: Geometric Construction}

\ArticleName{Toeplitz Quantization and Asymptotic
 Expansions:\\ Geometric Construction\footnote{This paper
is a contribution to the Special Issue on Deformation
Quantization. The full collection is available at
\href{http://www.emis.de/journals/SIGMA/Deformation_Quantization.html}{http://www.emis.de/journals/SIGMA/Deformation\_{}Quantization.html}}}

\Author{Miroslav ENGLI{\v S}~$^{\dag\ddag}$ and Harald UPMEIER~$^\S$}

\AuthorNameForHeading{M.~Engli{\v s} and H.~Upmeier}

\Address{$^\dag$~Mathematics Institute, Silesian University at Opava,\\
\hphantom{$^\dag$}~Na~Rybn\'\i\v cku~1, 74601~Opava, Czech Republic}

\Address{$^\ddag$~Mathematics Institute, \v Zitn\' a 25, 11567~Prague~1, Czech  Republic}
\EmailD{\href{mailto:englis@math.cas.cz}{englis@math.cas.cz}}

\Address{$^\S$~Fachbereich Mathematik, Universit\"at Marburg,
D-35032 Marburg, Germany}
\EmailD{\href{mailto:upmeier@mathematik.uni-marburg.de}{upmeier@mathematik.uni-marburg.de}}

\ArticleDates{Received October 01, 2008, in f\/inal form February 14,
2009; Published online February 20, 2009}

\Abstract{For a real symmetric domain $G_\R/K_\R$, with complexif\/ication
$G_\C/K_\C$, we introduce the concept of ``star-restriction'' (a~real
analogue of the ``star-products'' for quantization of K\"ahler manifolds)
and give a geometric construction of the $G_\R$-invariant dif\/ferential
operators yielding its asymptotic expansion.}


\Classification{32M15; 46E22; 47B35; 53D55}

\section{Introduction}
\label{SecINTRO}
Geometric quantization of (complex) K\"ahler manifolds is of particular
interest for \textit{symmetric} manifolds $B=G/K$ (of compact or non-compact
type). In~this case the Hilbert state space~$H$ carries an irreducible
representation of~$G$, whereas the various star products (Weyl calculus,
Toeplitz--Berezin calculus) describe the (associative) product of
\textit{observables} (operators on $H$) as an asymptotic  series of
$G$-invariant bi-dif\/ferential operators on~$B$.

In~this paper we introduce and study similar concepts for \textit{real}
symmetric manifolds (of f\/lat or non-compact type), emphasizing the interplay
between the real symmetric space and its ``hermitif\/ication'' which is a complex
hermitian space (of f\/lat or non-compact type). In~general, for a real-analytic
manifold $B_\R$ of dimension~$n$, a~complexif\/ication $B_\C$ is a complex
manifold of (complex) dimension~$n$, with $B_\R$ embedded (real-analytically)
as a totally real submanifold~\cite{AG,DHH,La}. If $B_\R = G_\R/K_\R$ is a symmetric space,
for a real (reductive) Lie group $G_\R$ with maximal compact subgroup $K_\R$,
we~write its hermitif\/ication as $B_\C = G_\C/K_\C$, where $G_\C$ denotes the
(real, semi-simple) biholomorphic isometry group and $K_\C$ is the maximal
compact subgroup. Thus, contrary to the usual notational conventions,
$G_\C$~is \textit{not} the complexif\/ication of $G_\R$ but the real Lie group
``in the complex setting''. For example, if $G_\R = SU(1,1)$ then $G_\C$
is given by $SU(1,1) \times SU(1,1)$ instead of $SL(2,\C)$; similarly,
for $G_\R=SO(1,1)$ we have $G_\C=SU(1,1)$.

On~the level of states, the interplay between a real symmetric space
$B_\R = G_\R/K_\R$ and its hermitif\/ication $B_\C = G_\C/K_\C$ corresponds
to a ``real-wave'' realization of $H_\C$ via a Segal--Bargmann transformation
\cite{Zhb}, which is invariant under the subgroup $G_\R\subset G_\C$.
On~the other hand, the real analogue of the star-product is not so obvious.
In~this paper (and its companion paper \cite{EU}) we introduce such a concept,
called ``star-restriction'' for real symmetric domains of non-compact type and
study its asymptotic expansion as a series of $G_\R$-invariant dif\/ferential
operators. Whereas the paper \cite{EU} establishes existence and uniqueness
of the asymptotic expansion, closely related to spectral theory and harmonic
analysis (spherical functions), the current paper gives a ``geometric
construction'' of the dif\/ferential operators involved, based on a
$G_\R$-invariant retraction $\pi : B_\C \to B_\R$.

We~emphasize that our $*$-restriction operator is a $G_\R$-equivariant map
\[
 \cC^\infty (B_\C) \to \cC^\infty  (B_\R)
\]
instead of a map $\cC^\infty(B_\R)\otimes\cC^\infty(B_\R)\to\cC^\infty(B_\R)$
analogous to the usual $*$-products. Thus we do not propose a quantization
method for general real symmetric domains (which may not be symplectic nor
even dimensional) but instead consider invariant operators which somewhat
resemble boundary restriction operators such as Szeg\"o or Poisson kernel
integrals. In~case $B_\R$ is the underlying real manifold of a complex
hermitian domain~$B$, then both concepts coincide and indeed yield the
well-known covariant quantization methods applied to the K\"ahler manifold~$B$.

In~order to illustrate the two concepts, consider the simplest non-f\/lat case of
the open unit disk $B\subset \C$ and its real form $B_\R = (-1,1)\subset \R$.
The~complexif\/ication $B^\C_\R$ coincides with~$B$, and we have a restriction
operator $\rho$, mapping a smooth function $f$ on $B=B^\C_\R$ to its
restriction~$\rho f$ on~$B_\R$. A~star-restriction is a \textit{deformation}
of the operator $\rho$, obtained by adding smooth, but non-holomorphic,
dif\/ferential operators on $B$ as higher order terms. In~the context of
symmetric domains, these dif\/ferential operators should be invariant under
the subgroup $G_\R$ of the holomorphic automorphism group $G$ of $B$ which
leaves $B_\R$ invariant.

Now consider instead the (usual) complex situation. Here $B$ is regarded as a
\textit{real} (symplectic) manifold, denoted by $B^\R$, whose complexif\/ication
$B^\R_\C$ is the product of $B$  and its complex conjugate $\overline B$, with
$B^\R$ embedded as the diagonal. Then a~star-product, regarded as a bilinear
operator acting on $f\otimes g$ (with $f$, $g$ smooth functions on $B$),
is~precisely a deformation of the usual product $f\cdot g$ by ($G$-invariant)
bi-dif\/ferential operators on $B$ or, equivalently, dif\/ferential operators on
$B^\R_\C = B\times\overline B$. Since $f\cdot g$ is nothing but the restriction
of $f\otimes g$ to the diagonal $B^\R \subset B^\R_\C$, we see that the concept
of star-restriction yields in fact the star-product for the special case where
the complexif\/ied domain is of product type. The~higher-dimensional case is
analogous.

In~order to state our main result concerning the asymptotic expansion (in~the
deformation parameter~$\nu$) of~a $*$-restriction operator as above, we~f\/irst
note that for the basic Toeplitz--Berezin calculus (the~only case considered in
detail here) the $*$-restriction operator is trivial for anti-holomorphic
functions so that we may concentrate on the holomorphic part, which is a~$G_\R$-covariant map
\[  \rho_\nu : \ \cO (B_\C) \to \cC^\infty (B_\R).  \]
Using deep facts from representation theory (of~the compact Lie groups $K_\R$
and $K_\C$), we~construct a family of dif\/ferential operators
\[  \rho^\bm : \ \cO (B_\C) \to \cC^\infty (B_\R)   \]
indexed by integer partitions $m_1 \geq \dots \geq m_r \geq 0$
(cf.~Def\/inition~\ref{d2.9}), and (in~Theorem~\ref{t2.12}) express~$\rho_\nu$
as an asymptotic series
\be
  \rho_\nu \sim \sum\limits_{\bm} \frac{1}{[\nu]_\bm}\ \rho^\bm,   \label{Ins1}
\ee
where the constants $[\nu]_\bm$ are generalized Pochhammer symbols.

Using the Fourier--Helgason transform on~$B_\R$, it~is conceivable
(see~\cite{EU} for the details) that $\rho_\nu$ can also be expressed
as an oscillatory integral
\be
 \rho_\nu F(x) \sim \int_{B_\C} F(z) \,a_\nu(z,x)\,e^{\nu S(z,x)} \,dz,
 \label{Ins2}
\ee
where $a_\nu$ is a suitable power series in~$\frac1\nu$, and the ``phase''
$S$ is a function on $B_\C\times B_\R$ invariant under the diagonal action
of~$G_\R$. This is reminiscent of the WKB-quantization programme of Karasev,
Weinstein and Zakrzewski \cite{Wei}, studied extensively in the context of
symplectic (i.e.~not necessarily Hermitian, or~even Riemannian) symmetric
spaces by Bieliavsky, Pevzner, Gutt, and other authors, see e.g.~\cite{BCG,Bie,BiP,BDS}. As~has already been pointed out above,
real symmetric domains need not be symplectic (in~fact, they can even be of
odd dimension), so~neither of the two approaches contains the other, and the
situations where they both apply include the original K\"ahler case of an
Hermitian symmetric space. A~thorough comparison of both methods~is, however,
beyond the scope of this paper. For~the f\/lat cases of $B_\R=\R^d$ and $B_\R=
\C^d$, the~expansions (\ref{Ins2}) were obtained quite explicitly, and for
a whole one-parameter class of calculi which includes the Toeplitz calculus,
by~Arazy and the second author~\cite{AUweyl}.

In~Section~\ref{SecDOUBLE} the~asymptotic series (\ref{Ins1}) are computed for the simplest
cases of (real or complex) dimension~1. In~general, f\/inding explicit formulas
may be  quite dif\/f\/icult, but there is some hope that at least all symmetric
domains  of rank~1 (i.e., hyperbolic spaces in $\R^n$, $\C^n$, $\qH^n$ and the
Cayley plane) can be treated in a unif\/ied and explicit way.

\section{Preliminaries}
\label{SecPRELIM}

One~of the most inspiring examples of deformation quantization is the
Berezin quantiza\-tion~\cite{BerCMP,BerQ} using the Berezin transform
and Toeplitz operators (originally called co- and contra-variant symbols,
respectively). Although it has subsequently been generalized and extended
to various classes of compact and noncompact K\"ahler manifolds \cite{BMS,E32,KS,RT}, the theory is still richest in its original
setting of complex symmetric spaces, or~bounded symmetric domains, in~$\C^d$
\cite{BerQCO}, due~to the powerful machinery available from Lie groups and
their representation theory on the one hand \cite{Helg,U1}, and from
the theory of Jordan triple systems on the other~\cite{Loo}.

More specif\/ically, let $B=G/K$ be an irreducible  bounded symmetric domain
in $\C^d$ in the Harish-Chandra realization, with $G$ the identity connected
component of the group of all biholomorphic self-maps of~$B$ and $K$~the
stabilizer of the origin. For~$\nu>p-1$, $p$~being the genus of~$B$,
let~$H^2_\nu(B)$ denote the standard weighted Bergman space on~$B$,
i.e.~the subspace of all holomorphic functions in $L^2(B,d\mu_\nu)$, with
\[  d\mu_\nu(z) = c_\nu \, K(z,z)^{1-\nu/p}\,dz ,   \]
where $dz$ stands for the Lebesgue measure, $K(z,w)$ is the ordinary
(unweighted) Bergman kernel of~$B$, and $c_\nu$ is a normalizing constant
to make $d\mu_\nu$ a probability measure. The~space $H^2_\nu(B)$ carries
the unitary representation $\unu$ of $G$ given~by
\[  \unu_g f(z) = f(g^{-1}(z)) \cdot J_{g^{-1}}(z)^{\nu/p},
\qquad g\in G, \, f\in H^2_\nu(B),   \]
where $J_g$ denotes the complex Jacobian of the mapping~$g$. (In~general,
if~$\nu/p$ is not an integer, then $\unu$ is only a projective representation
due to the ambiguity in the choice of the power $J_{g^{-1}}(z)^{\nu/p}$.)

By~a~\emph{covariant operator calculus}, or~\emph{covariant quantization},
on~$B$ one understands a mapping $\AA:f\mapsto\AA_f$ from functions on $B$
into operators on $H^2_\nu(B)$ which is $G$-covariant in the sense that
\[  \AA_{f\circ g} = \unu_g{}^*  \AA_f  \unu_g,  \qquad\forall \, g\in G.   \]
In~most cases, such calculi can be built by the recipe
\[  \AA_f = \int_B f(\zeta)   \AA_\zeta \, d\mu_0(\zeta)   \]
where $d\mu_0$ is a $G$-invariant measure on~$B$, and $\AA_\zeta$ is a family
of operators on $H^2_\nu(B)$ labelled by $\zeta\in B$ such that
\[  \AA_{g(\zeta)} = \unu_g \AA_\zeta \unu_g{}^*,  \qquad\forall \, g\in G.  \]
(One~calls such a family a \emph{covariant operator field} on~$B$. One~also
usually normalizes the measure $d\mu_0$ so that $\AA_\jedna$ is the identity
operator.) Note that in view of the transitivity of the action of $G$ on~$B$,
any~covariant operator f\/ield is uniquely determined by its value $\AA_0$ at
the origin~$\zeta=0$.

The~best known examples of such calculi are the \textit{Toeplitz calculus}
$\TT$ and the \textit{Weyl calculus}~$\WW$, corresponding to
$\TT_0=(\cdot|\jedna)\jedna$ (the~projection onto the constants) and
$\WW_0f(z)=f(-z)$ (the~ref\/lection operator), respectively.

In~addition to bounded symmetric domains, we~will also consider the
\textit{complex flat} case of a~Hermitian vector space $B=Z\approx\C^d$,
with $B=G/K$ for $G$ the group of all orientation-preserving rigid motions
of~$Z$, and $K = U(Z) \approx U_d (\C)$ the stabilizer of the origin in~$G$;
the~spaces~$H^2_\nu(Z)$ will then be the Segal--Bargmann spaces of all entire
functions which are square-integrable with respect to the Gaussian measure
\[  d\mu_\nu(z) = \left(\frac\nu\pi\right)^d  \; e^{-\nu\|z\|^2}\,dz,   \]
and $\unu$ will be the usual Schr\"odinger representation. In~this setting,
the Weyl calculus $\WW$ above is just the well-known Weyl calculus from the
theory of pseudodif\/ferential operators \cite{Foll}.

Given a covariant operator calculus~$\AA$, the associated \emph{star product}
$*$ on functions on $B$ is def\/ined~by
\be
  \AA_{f*g} = \AA_f \AA_g.         \label{e:tTF}
\ee
It~follows from the construction that the star-product is \emph{$G$-invariant}
in the sense that
\be   \label{GIN}
 (f\circ\phi)*(g\circ\phi) = (f*g)\circ\phi \qquad \forall \,\phi\in G.
\ee
While $f*g$ is a well-def\/ined object for some calculi (e.g.~for $\AA=\WW$,
at~least on $\C^d$ and rank one symmetric domains, see~\cite{AUweyl}),
in~most cases (e.g.~for $\AA=\TT$, the~Toeplitz calculus), it makes sense
only for very special functions~$f$,~$g$ and (\ref{e:tTF}) is then usually
understood as an equality of asymptotic expansions as the Wallach
parameter $\nu$ tends to inf\/inity. For~instance, for $\AA=\TT$, it~was
shown~in~\cite{BLU} that for any $f,g\in\cC^\infty(B)$ with compact support,
\[  \| \TT_f \TT_g - \TT_{\sum_{j=0}^N \nu^{-j} C_j(f,g)} \| = O\big(\nu^{-N-1}\big) \]
as $\nu\to\infty$, for some bilinear dif\/ferential operators $C_j$
(not~depending on~$f$,~$g$ and $\nu$). (The~assumption of compact support
can be relaxed~\cite{E}.) We~can thus def\/ine $f*g$ as the formal
power series
\[ f*g := \sum_{j=0}^\infty \nu^{-j} \, C_j(f,g).   \]
Interpreting $\nu$ as the reciprocal of the Planck constant, we~recover the
Berezin--Toeplitz star product~\cite{Schli}, which is the dual to Berezin's
original star-product mentioned above~\cite{E32}. (This~approach to the Berezin
and Berezin--Toeplitz star-products, i.e.~using covariant operator calculi and
the def\/inition~(\ref{e:tTF}), is~not the traditional way of constructing the
$G$-invariant Berezin quantization, however, for~the case of bounded symmetric
domains these two are equivalent~\cite{E41}.)

Viewing the Planck parameter $\nu$ as f\/ixed for the moment, the formula
(\ref{GIN}) means that one can view $*$ as a mapping from the tensor product
\[ * : \ \cC^\infty(B\times B)\cong\cC^\infty(B)\otimes\cC^\infty(B) \to
 \cC^\infty(B), \qquad f\otimes g\to f*g,   \]
which intertwines the $G$-action $f\mapsto f\circ\phi$, $\phi\in G$,
on~$\cC^\infty(B)$ with the diagonal $G$-action $f\otimes g\mapsto(f\circ\phi)
\otimes(g\circ\phi)$ of $G$ on $\cC^\infty(B\times B)$. This observation can
be used as a starting point for extending the whole quantization procedure
also to \emph{real} bounded symmetric domains $B_\R\subset\R^d$, as~follows.

Suppose $Z_\C$ is an \textit{irreducible} hermitian Jordan triple \cite{Loo,U1} endowed with a (conjugate-linear) involution
\[  z \mapsto z^\#   \]
which preserves the Jordan triple product and therefore the unit ball $B_\C$
of $Z_\C$, i.e.~$(B_\C)^\#=B_\C$. Def\/ine the real forms
\begin{gather*}
 Z_\R  := \{z\in Z_\C : z^\# = z\}, \\
 B_\R  := \{z\in B_\C : z^\# = z\} = Z\cap B_\C.
\end{gather*}
For~the groups $G_\C:=\Aut(B_\C)$, $K_\C:=\Aut(Z_\C)$ we have the subgroups
\begin{gather*}
 G_\R  := \{g \in G_\C : g(z^\#) = g(z)^\#\} , \\
 K_\R  := \{k \in K_\C : k z^\# = (kz)^\#\} = G_\R \cap K_\C
 \end{gather*}
acting on $B_\R$ and~$Z_\R$, respectively. In~this situation $Z_\R$
is an irreducible real Jordan triple, $G_\R$~is a reductive Lie group
(it~may have a nontrivial center), and
\[ B_\R = G_\R/K_\R    \]
is~an irreducible real bounded symmetric domain. Up~to a few exceptions,
all non-hermitian Riemannian symmetric spaces of non-compact type arise
in this way \cite[Chapter~11]{Loo}.

A~\emph{covariant quantization} (or~\emph{covariant extension}) on~the
real bounded symmetric domain $B_\R$ is a map $f\mapsto\AA_f$ from
$\cC^\infty(B_\R)$ into $H^2_\nu(B_\C)$ such that
\[  \AA_{f\circ g} = \unu_g{}^* \AA_f    \]
for all $g\in G_\R$. The~counterpart of the star product, associated to a
covariant quantization $\AA$ on $B_\R$ and a covariant quantization $\AAC$
on~$B_\C$, is~the \emph{star restriction}
\[ \rho=\rho_\nu: \  \cC^\infty(B_\C) \to \cC^\infty(B_\R)  \]
def\/ined~by
\be
  \AA_{\rho F} = \AAC_F I,  \label{e:tTI}
\ee
where
\[  I(z) = I_\nu(z) = \knu(z,z^\#)^{1/2}   \]
is~the unique $G_\R$-invariant holomorphic function on~$B_\C$ satisfying
$I(0)=1$. In~addition, we~will again consider the above construction also
in the case of the Segal--Bargmann spaces for an involutive Hermitian vector
space $Z_\C\approx\C^d$, with the ordinary complex conjugation as the
involution $z\mapsto z^\#$; thus $B=Z_\R\approx\R^d$.

In~most cases, covariant extensions can again be constructed by the recipe
\[  \AA_f = \int_{B_\R} f(\zeta)  \AA_\zeta \,d\mu_0(\zeta),    \]
where $d\mu_0$ is the $G_\R$-invariant measure in~$B_\R$, and $\AA_\zeta$ is a
family of holomorphic functions (not necessarily belonging to $H^2_\nu(B_\C)$)
labelled by $\zeta\in B_\R$ which is covariant in the sense that
\[  \AA_{g(\zeta)} = \unu_g \AA_\zeta, \qquad \forall\,  g\in G_\R,\,
 \zeta\in B_\R.   \]
As~before, one~usually normalizes $d\mu_0$ so that $\AA_\jedna=I$.
The~prime example is now the \textit{real Toeplitz calculus} $\AA=\TT$
corresponding to~$\AA_0=\jedna$ (the~function constant~one)
\cite{Z, N,vDP,AUlink,AUc}; there is also a
notion of \textit{real Weyl calculus}, but~it is more complicated~\cite{AU2}.

Here is how the complex hermitian case of a bounded symmetric domain
$B\subset\C^d$ from the beginning of this section can be recovered within
the more general real framework. Identify $B$ with the ``diagonal'' domain
\[  B^\R := \{(z,\oz) : z\in B\} \subset Z^\R  := \{(z,\oz) : z \in Z\},  \]
where the bar indicates that we consider the ``conjugate'' complex structure
for the second component. The complexif\/ications
\begin{gather*}
B_\C^\R   = \{(z,\ow) : z,w\in B\} = B\times \overline B, \\
Z_\C^\R   = \{(z,\ow) : z,w\in Z\} = Z\times \overline Z
\end{gather*}
are endowed with the f\/lip involution
\[  (z,\ow)^\# := (w,\oz)   \]
having f\/ixed points $B^\R$ and $Z^\R$, respectively. Similarly we can identify
$G$, $K$ with the groups
\begin{gather*}
 G^\R  := \{(g,\overline g) : g\in G\}, \\
 K^\R  := \{(k,\overline k) : k\in K\}
\end{gather*}
which act ``diagonally'' on $B^\R$ and $Z^\R$, respectively, and whose
complexif\/ications
\begin{gather*}
 G_\C  := \{(g_1,\overline g_2) : g_1, g_2\in G\} = G \times \overline G, \\
 K_\C  := \{(k_1,\overline k_2) : k_1, k_2\in K\} = K \times \overline K
\end{gather*}
act~on~$B_\C$ and~$Z_\C$, with $B_\C=G_\C/K_\C$.

Since $H^2_\nu(B)$ is a reproducing kernel space (with reproducing kernel
$\knu(x,y)=h(x,y)^{-\nu}$, where $h(x,y)=[K(x,y)/c_p]^{-1/p}$ is the Jordan
determinant polynomial), any~bounded linear operator on $H^2_\nu(B)$ is
automatically an integral operator: namely,
\[ T f(z) = \int_B f(w)   \wt T(z,w) \, d\mu_\nu(w),   \]
with
\[ \wt T(z,w) = \overline{(T^*\knu(\cdot,z))(w)}
 = (T\knu(\cdot,w)|\knu(\cdot,z)) .    \]
This follows from the identity
$ Tf(z) = (Tf|\knu(\cdot,z)) = (f|T^*\knu(\cdot,z)).  $
In~this way, we~may identify operators on $H^2_\nu(B)$ with (some) functions
on~$B\times B$, holomorphic in the f\/irst and anti-holomorphic in the second
variable; that~is, with holomorphic functions on~$B_\C$.
Upon~this identif\/ication, the~covariant quantization rule
$f\mapsto\AA_f$ becomes simply a (densely def\/ined) operator $f\mapsto\wt\AA_f$
from $\cC^\infty(B_\R)$ into the Hilbert space
\[ H^2_\nu(B_\C) \approx H^2_\nu(B)\otimes \overline{H^2_\nu(B)}  \]
corresponding to the Hilbert--Schmidt operators, and the covariance condition
means that $\wt\AA$ is equivariant under $G_\R\approx G$, i.e.~intertwines the
$G$-action on the former with the diagonal $G$-action on the latter:
\[  \wt\AA_{f\circ g} = \big(\unu_g{}^* \otimes
 \overline{\unu_g{}^*} \big)   \wt\AA_f.   \]
Similarly, upon taking $\AAC=\AA\otimes\AA$, and identifying pairs $f$, $g$
of functions on~$B$ with the function $F(x,y)=f(x)\overline{g(y)}$
on~$B_\C$, (\ref{e:tTI})~reduces just to~(\ref{e:tTF}).
Note, however, that the complexif\/ied domain $B_\C$ is now no longer
irreducible, but of ``product type''. We~will henceforth refer to this
situation, i.e.~of $B_\R=B$, $B_\C=B\times\overline B$ with a bounded
symmetric domain $B\subset\C^d$, as the ``complex''~case.

To~each covariant extension (or~quantization) $\AA$ we can consider its adjoint
$\AA^*$ from $H^2_\nu(B_\C)$ into functions on~$B_\R$, def\/ined with respect to
the inner products in $H^2_\nu(B_\C)$ and $L^2(B_\R,d\mu_0)$. That~is,
\be
 (\AA^* f|\phi) _{L^2} = (f|\AA_\phi)_\nu ,   \qquad \forall\,
 \phi\in L^2(B_\R,d\mu_0), \quad \forall \,f\in H^2_\nu(B_\C).    \label{e:DUAL}
\ee
One~sometimes calls $\AA^*$ a \emph{covariant restriction}; this should not
be confused with the star-restriction~$\rho$, which is a map from $\cC^\infty
(B_\C)$ into functions on~$B_\R$.

One~can also consider the associated \emph{link transform}, which is the
composition $\AA^*\AA$, a~$G_\R$-invariant operator on functions on~$B_\R$.
In~particular, for the Toeplitz calculus $\AA=\TT$, the~link transform
\[   \TT^*\TT =: \cB_\nu    \]
is~the \emph{Berezin transform}, introduced for the ´`complex`''  case in
Berezin's original papers (cf.~Section~\ref{SecDOUBLE} below).

A~crucial role in the analysis on complex bounded symmetric domains is played
by the Peter--Weyl decomposition of holomorphic functions on $B$ under the
composition action $f\mapsto f\circ k$ of the (compact) group~$K$. Namely,
the vector space $\cP$ of all holomorphic polynomials on $\C^d$ decomposes
under this action into non-equivalent irreducible components
\[ \cP = \sum_\bm \cP_\bm  \]
labelled by \emph{partitions} (or~\emph{signatures}) $\bm\in\N^r_+$, that~is,
by~$r$-tuples of integers $m_1\ge m_2\ge\dots\ge m_r\ge0$, where $r$ is the
rank of~$B$. With respect to the Fock inner product
\[ (p|q)_F := \int_{\C^d} p(z)   \overline{q(z)}   e^{-|z|^2} \, dz
 = p(\partial) q^*(0), \qquad q^*(z):=\overline{q(z^\#)},   \]
each Peter--Weyl space $\cP_\bm$ possesses a reproducing kernel $K_\bm(z,w)$,
$z,w\in Z$. It~was shown by Arazy and {\O}rsted \cite{AO} that the Berezin
transform~$\cB_\nu$ admits the asymptotic expansion
\[
 \cB_\nu = \sum_\bm \frac{\cE_\bm}{(\nu)_\bm}
  \qquad\text{as} \quad \nu\to+\infty,   
\]
where $\cE_\bm$ is the $G$-invariant dif\/ferential operator on $B$ determined
(uniquely) by the requirement that
\[ \cE_\bm f(0) = K_\bm(\partial,\partial) f(0),
 \qquad \forall \, f\in\cC^\infty(B);    \]
while $(\nu)_\bm$ is the multi-Pochhammer symbol
\[ (\nu)_\bm = \prod_{j=1}^r \frac{\Gamma(\nu-\frac a2(j-1) + m_j)}
 {\Gamma(\nu-\frac a2(j-1) )} ,   \]
$a$~being the so-called \emph{characteristic multiplicity} of~$B$. Analogously,
it~was shown in \cite{E} that  the star-product (\ref{e:tTF}) arising from
the Toeplitz calculus $\AA=\TT$ admits an expansion
\be
  f*g = \sum_\bm \frac{A_\bm(f,g)}{(\nu)_\bm},   \label{e:AM}
\ee
where $A_\bm$ are certain (rather complicated) $G$-invariant (cf.~(\ref{GIN}))
bi-dif\/ferential operators.

The~main purpose of the present paper is an extension of the last formula
to real symmetric domains. That~is, to obtain a decomposition of the star
restriction operator
\be
 \rho_\nu = \sum_\bm \frac{\rho^\bm}{[\nu]_\bm}  \label{MAIN}
\ee
with some $G_\R$-invariant dif\/ferential operators $\rho^\bm:\cC^\infty(B_\C)
\to\cC^\infty(B_\R)$ (independent of~$\nu$) and generalized  ``Pochhammer
symbols'' $[\nu]_\bm$.

A~prominent role in our analysis is  played by holomorphic
polynomials on $Z_\C$ which are invariant under the group $K_\R$. In~the
Peter--Weyl decomposition under $K_\C$ mentioned above, partitions $\bn$
for which $\cP_\bn$ contains a nonzero $K_\R$-invariant vector are
called ``even'', and are in one-to-one correspondence with
partitions $\bm$ of length $r_\R=\operatorname{rank}B_\R$; furthermore,
for~each ``even'' Peter--Weyl space the subspace of $K_\R$-invariant vectors
is one dimensional, consisting only of multiples of a certain
polynomial which (under an appropriate normalization) we denote by~$E^\bm$.
For~more details, including the description of~$E^\bm$, bibliographic
references, etc., as~well as for the various preliminaries and notation
not introduced here, we~refer to \cite{Z,EU}.

The~construction of the decomposition (\ref{MAIN}) for general real symmetric
domains is carried out in Section~\ref{SecMOYAL}. In~Section~\ref{SecDOUBLE}
it is shown that the decomposition obtained indeed reduces to~(\ref{e:AM})
for the ``complex'' case. The final  Section~\ref{SecEXMPS}
contains a few examples with more or less explicit formulas for $\rho^\bm$
and~$[\nu]_\bm$. For the reader's convenience, we~are also attaching
 a table of all real bounded symmetric domains and their
various parameters.

In~some sense, our results can be perceived as a step towards
building a version of Berezin's quantization for real (as~opposed to K\"ahler)
manifolds as phase spaces.

\section{Invariant retractions and Moyal restrictions}
\label{SecMOYAL}
As a f\/irst step towards a geometric construction of asymptotic expansions for
the Moyal type restriction, we obtain an integral representation for the Moyal
restriction operator, def\/ined in terms of $G_\R$-invariant retractions
\[
 \pi : \ B_\C \to B_\R.
\]
Here $B_\R$ is an irreducible real symmetric domain of rank $r$, in~its bounded
realization (real Cartan domain) and $B_\C$ is the open unit ball of the
complexif\/ication $Z_\C$, which is a complex hermitian bounded symmetric domain,
not necessarily irreducible \cite{Loo,FKc,U1,Hu}.

We~will  assume that the preimage $\pi^{-1}(0)$ of the origin
$0\in B_\R$ has the form
\be
  \pi^{-1}(0) = B_\C \cap Y = \Lambda B_\R    \label{e:YY}
\ee
for some real vector subspace $Y\subset B_\C$ and real-linear $K_\R$-invariant
map $\Lambda: Z_\R \to Z_\C$. Our~construction in fact works even without these
assumptions (cf.~Remark~\ref{RemGe} below), but~all situations studied in this
paper will be of the above form.

Let $h_\C : Z_\C \times \overline{Z}_\C\to \C$ denote the Jordan triple
determinant (cf.~\cite{Loo}) of $Z_\C$ and def\/ine the \textit{Berezin kernel}
\[
\cB_\nu : \ B_\C \to \C
\]
by
\begin{equation}\label{e1.1}
 \cB_\nu (z) := h_\C(z,z)^\nu/|h_\C(z,z^\sharp)|^\nu,
\end{equation}
where $z \mapsto z^\sharp$ is the involution with real form $B_\R$. Note that
$h_\C(z,w) \neq 0$ for all $z,w\in B_\C$.

\begin{proposition}
The~Berezin kernel $\cB_\nu$ is $G_\R$-invariant, i.e.,
\[
\cB_\nu (gz) = \cB_\nu (z)
\]
for all $g\in G_\R$ and $z\in B_\C$.
\end{proposition}

\begin{proof}
Since $G_\R\subset G_\C$ we have
\[
 h_\C(gz, gw)^\nu = j_\nu (g,z)\ h_\C (z,w)^\nu\,  \overline{j_\nu (g,w)}
\]
for all $z,w \in B_\C$, where
\[
 j_\nu (g,z) = [\det\, g'(z)]^{\nu/p}
\]
and $p$ is the (complex) genus of $B_\C$. For $g\in G_\R$ we have
\[
  g(z)^\sharp = g(z^\sharp)
\]
and
\[
 \overline{j_\nu (g,z)} = j_\nu (g,z^\sharp)
\]
since these relations are anti-holomorphic in $z\in B_\C$ and hold for
$z=z^\sharp$. It~follows that
\begin{gather*}
    \frac{h_\C(gz, gz)^\nu}{|h_\C (gz, (gz)^\sharp)|^\nu} =
    \frac{h_\C (gz,gz)^\nu}{h_\C (gz, g(z^\sharp))^{\nu/2}\,
    h_\C (g(z^\sharp),gz)^{\nu/2}} \\[1em]
 \phantom{\frac{h_\C(gz, gz)^\nu}{|h_\C (gz, (gz)^\sharp)|^\nu}}{} =  \frac{j_\nu (g,z)\ h_\C (z,z)^\nu\ \overline{j_\nu (g,z)}}{j_{\nu/2}
   (g,z)\ h_\C (z,z^\sharp)^{\nu/2}\ \overline{j_{\nu/2} (g,z^\sharp)}\
   j_{\nu/2} (g,z^\sharp)\ h_\C (z^\sharp, z)^{\nu/2}\ \overline{j_{\nu/2}
   (g,z)}} \\[1em]
\phantom{\frac{h_\C(gz, gz)^\nu}{|h_\C (gz, (gz)^\sharp)|^\nu}}{} = \frac{h_\C  (z,z)^\nu}{|h_\C (z,z^\sharp)|^\nu}\ \frac{j_{\nu/2} (g,z)\
   \overline{j_{\nu/2} (g,z)}}{\overline{j_{\nu/2} (g,z^\sharp)}\ j_{\nu/2}
   (g,z^\sharp)} = \frac{h_\C  (z,z)^\nu}{|h_\C (z,z^\sharp)|^\nu}.
\end{gather*}
Another proof can be given by observing that, using the familiar
transformation rule for~$h_\C$,
\[ \frac{h_\C(gz,gz)}{|h_\C(gz,gz^\#)|}
 = \frac {\dfrac{h_\C(z,z) h_\C(a,a)} {|h_\C(z,a)|^2}}
         {\Big|\dfrac{h_\C(z,z^\#) h_\C(a,a)} {h_\C(z,a) h_\C(a,z^\#)}\Big|}
 = \frac{h_\C(z,z)}{|h_\C(z,z^\#)|} \, \Big|\frac{h_\C(a,z)}{h_\C(a,z^\#)},  \]
where $g\in G_\C$ and $a=g^{-1}(0)$. If~$g\in G_\R$, then $gz^\#=(gz)^\#$,
while $h_\C(a,z^\#)=\overline{h_\C(a^\#,z)}=\overline{h_\C(a,z)}$ (as~$a^\#=a$)
whence $|h_\C(a,z^\#)|=|h_\C(a,z)|$. Thus $\cB_\nu(gz)=\cB_\nu(z)$.
\end{proof}
The~relationship between the Moyal restriction operator
\[
 \rho_\nu : \ \cC^\infty (B_\C) \to \cC^\infty (B_\R)
\]
and the Berezin kernel $\cB_\nu$ is given by the following
result.

\begin{proposition}\label{p1.2}
For $G\in \cO (B_\C)$ and $F \in \cC^\infty (B_\C)$ we have, if~{ }$\nu$ is
large enough,
\[
 \int_{B_\R}  dx\, h_\C(x,x)^{\frac{\nu-p}{2}}
  \overline{G (x)}  (\rho_\nu\, F)(x)
 = \int_{B_\C}  dz\, h_\C (z,z)^{-p}  \cB_\nu (z)
  \overline{(G/I_\nu)(z)} F(z),
\]
where
\[
 I_\nu (z) = h_\C(z,z^\sharp)^{-\nu/2}.
\]
\end{proposition}

\begin{proof}
The~Toeplitz restriction map $\cT^*_\R$ satisf\/ies
\[
 (\cT^*_\R\, G)(x) = h_\C(x,x)^{\nu/2}\, G(x) = (G/I_\nu)(x)
\]
for all $x\in B_\R$ \cite{Z,AUlink}. Using the duality relation
(\ref{e:DUAL}) and the def\/inition (\ref{e:tTI}) of $\rho_\nu$ we obtain
\begin{gather*}
  \int_{B_\R} dx\, h_\C (x,x)^{\frac{\nu-p}{2}}  \overline{G(x)}
   \ (\rho_\nu\, F)(x)
   =  \int_{B_\R}  dx\, h_\C (x,x)^{-p/2}  \overline{(G/I_\nu) (x)}
   (\rho_\nu F)(x) \\
\qquad{} = \int_{B_\R} dx\, h_\C (x,x)^{-p/2}  \overline{(\cT^*_\R G)(x)}
   (\rho_\nu F)(x)
   = (\cT^*_\R G|\rho_\nu F)_{B_\R} = (G|\cT_\R\, \rho_\nu F)_\nu \\
\qquad{} = (G|\cT_\C (F)\, I_\nu)_\nu =  (G|F\cdot I_\nu)_\nu
   = \int_{B_\C} dz \, h_\C (z,z)^{\nu-p}  \overline{G(z)}  F(z)
       I_\nu (z) \\
\qquad{} = \int_{B_\C} dz \, h_\C (z,z)^{\nu-p}   \overline{(G/I_\nu)(z)}
     F(z)  |I_\nu(z)|^2 .
\end{gather*}
Since
\[
 h_\C (z,z)^\nu  |I_\nu (z)|^2 = \cB_\nu (z)
\]
the assertion follows.
\end{proof}

\begin{corollary}
For $G\in \cO (B_\C)$ and $F \in \cC^\infty (B_\C)$, we~have
$\rho_\nu(\overline G\, F) = \overline G \; \rho_\nu F$.
\end{corollary}

It~follows from~(\ref{e:YY}) that $Y\subset Z_\C$ is a $K_\R$-invariant
subspace such that
\[
 Z_\C = Z_\R \oplus Y
\]
(direct sum of real vector spaces). For $x\in B_\R$, let $\gamma_x \in G_\R$
be the ``transvection'' sending $0$ to~$x$, explicitly given~by
\begin{equation*}
 \gamma_x (y) = x + B(x,x)^{1/2}  (y^{-x}),
\end{equation*}
where $B$ is the Bergman operator and
\begin{equation*}
 y^x = B(y,x)^{-1} (y-Q_y x)
\end{equation*}
is the so-called quasi-inverse~\cite{Loo}.

\begin{lemma}  \label{lemmPhi}
The~mapping $\Phi : B_\R \times (Y\cap B_\C) \to B_\C$ defined by
\[
 \Phi (x,y) = \gamma_x (y)\qquad (x\in B_\R, \ \ y\in Y\cap B_\C)
\]
is a real-analytic isomorphism, whose derivative at $(0,y)$ is given by
\[
 \Phi'(0,y)(\xi,\eta) = \xi + \eta - \{y\xi y\}
\]
for all $\xi \in Z_\R = T_x (B_\R)$, $\eta \in Y = T_y (Y\cap B_\C)$.
\end{lemma}

\begin{proof} For $z\in B_\C$, set $x:=\pi z$ and $y=\gamma_{-x}z$
$(=\gamma_x^{-1}z)$. Then $x\in B_\R$ while, by~the $G_\R$-invariance of~$\pi$,
\[ \pi y = \gamma_{-x} \pi z = \gamma_{-x} x = 0,  \]
so $y\in Y\cap B_\C$. This proves that $\Phi$ is surjective.
Similarly, if~$\Phi(x,y)=\Phi(x',y')$ for some $x,x'\in B_\R$ and $y,y'\in Y
\cap B_\C$, then $x=\gamma_x 0=\gamma_x\pi y=\pi\Phi(x,y)=\pi\Phi(x',y')=x'$
and $y=\gamma_{-x}\Phi(x,y)=\gamma_{-x'}\Phi(x',y')=y'$, showing that $\Phi$
is injective. It~remains to prove the formula for the derivative. For this,
we~will use some of the formulas collected in \cite[Appendix~A1--A3]{Loo}.
For the quasi-inverse
\[
 \Psi (x,y) = x^y
\]
we obtain, by def\/inition,
\[
 \Psi (\xi, y) = B(\xi, y)^{-1} (\xi - Q_\xi\, y)
\]
and hence
\[
 (\partial_1\, \Psi)(0,y)\, \xi = \xi.
\]
Using the \textit{symmetry formula} \cite[A3]{Loo} we obtain
\[
  \Psi (x,\eta) = x^\eta = x + Q_x (\eta^x)
  = x + Q_x\ B(\eta,x)^{-1}  (\eta - Q_\eta\, x)
\]
and hence
\[
 (\partial_2  \Psi)(x,0)  \eta = Q_x  \eta.
\]
Now the \textit{addition formulas} \cite[A3]{Loo} yield
\[
 (x+\xi)^y = x^y + B(x,y)^{-1}(\xi^{(y^x)})
\]
and hence
\[
 (\partial_1  \Psi)(x,y)  \xi = B(x,y)^{-1}  (\partial_1  \Psi)(0,y^x)
 \xi = B(x,y)^{-1}  \xi.
\]
Similarly, we have
\[
 x^{(y+\eta)} = (x^y)^{\eta}
\]
and hence, with (JP28) from~\cite[A2]{Loo},
\[
 (\partial_2  \Psi)(x,y)  \eta = (\partial_2  \Psi)(x^y, 0)  \eta
 = Q_{x^y}  \eta = B(x,y)^{-1}  Q_x  \eta.
\]
It~follows that
\[
 \Psi' (x,y)(\xi, \eta) = B(x,y)^{-1} (\xi + Q_x  \eta).
\]
Since $B(x,x)^{1/2}$ is an even function of $x$, its derivative at $x=0$
vanishes and we obtain for
\[
 \Phi (x,y) = \gamma_x (y)  =  x + B(x,x)^{1/2}  y^{-x}
 =  x + B (x,x)^{1/2}  \Psi (y,-x)
\]
the derivatives
\[
 \partial_1  \Phi (0,y)  \xi = \xi - \partial_2  \Psi (y,0)  \xi
 = \xi - Q_y  \xi
\]
and
\[
 \partial_2  \Phi (0,y)  \eta =  \partial_1  \Psi (y,0)  \eta =
 B(y,0)^{-1}  \eta = \eta.
\]
Therefore
\begin{gather*}
 \Phi' (0,y)(\xi, \eta) = (\partial_1  \Phi) (0,y)  \xi + (\partial_2  \Phi)
 (0,y)  \eta = \xi + \eta - Q_y  \xi.\tag*{\qed}
\end{gather*}\renewcommand{\qed}{}
\end{proof}

\begin{corollary} \label{coro1}
For all $y\in Y \cap B_\C$ we have
$
 \det  \Phi'(0,y) = {\det}_{Z_\R} (I-Q_y).
$
\end{corollary}

Def\/ine $\ps_\nu : \cC^\infty (B_\C) \to \cC^\infty (B_\R)$ by
\begin{gather*}
   (\ps_\nu F)(x) := h_\C (x,x)^{p/2} \int_{Y\cap B_\C} dy\,
     F(\gamma_x\,y)  |\det  \Phi' (x,y)|\cdot
   h_\C (\gamma_x  y, \gamma_x  y)^{-p}  \cB_\nu (y) \\
 \phantom{(\ps_\nu F)(x)}{} = h_\C (x,x)^{p/2} \int_{Y\cap B_\C} dy\, F(\gamma_x\, y)
   |\det  \Phi' (x,y)|  h_\C (\gamma_x y, \gamma_x y)^{\nu-p}
   |h_ \C (\gamma_x  y, (\gamma_x  y)^\sharp)|^{-\nu}
\end{gather*}
for all $F\in \cC^\infty (B_\C)$ and $x\in B_\R$. Here $\Phi'(x,y)$
is the derivative of $\Phi$ at $(x,y)\in B_\R \times  (Y \cap B_\C)$.
If~$f\in\cC^\infty (B_\R)$, then $f\circ \pi \in \cC^\infty (B_\C)$ and
\[
 (f\circ \pi)(\gamma_x y) = f(\gamma_x \pi(y)) = f(\gamma_x 0) = f(x).
\]
It~follows that
\begin{equation}\label{e1.3}
 \ps_\nu ((f\circ \pi)  F) = f\cdot (\ps_\nu F),
\end{equation}
i.e.~$\ps_\nu$ behaves like a ``conditional'' expectation.

\begin{proposition}\label{p1.5}
For $F \in \cC^\infty (B_\C)$ we have
\[
 \int_{B_\R} dx\, h_\C (x,x)^{-p/2}  (\ps_\nu F)(x)
 = \int_{B_\C} dz\, h_\C (z,z)^{-p}  \cB_\nu (z)  F(z).
\]
\end{proposition}

\begin{proof}
The~change of variables $z = \gamma_x (y) = \Phi(x,y)$ yields in view of the
invariance of $\cB_\nu$
\begin{gather*}
   \int_{B_\C} dz\, h_\C (z,z)^{-p}  \cB_\nu (z)  F(z)   =  \int_{B_\R} dx \int_{Y\cap B_\C} dy\, |\det
    \Phi' (x,y)|  h_ \C (\gamma_x\, y, \gamma_x  y)^{-p}  \cB_\nu (y)
      F(\gamma_x  y) \\
\phantom{\int_{B_\C} dz\, h_\C (z,z)^{-p}  \cB_\nu (z)  F(z) }{} = \int_{B_\R} dx\, h_\C(x,x)^{-p/2}  (\ps_\nu  F)(x).\tag*{\qed}
\end{gather*}\renewcommand{\qed}{}
\end{proof}

\begin{corollary}
The~operator $\ps_\nu$ is $G_\R$-invariant, i.e., we have
\[
 \ps_\nu (F\circ g) = (\ps_\nu F)\circ g
\]
for all $F\in \cC^\infty  (B_\C)$ and $g\in G_\R$.
\end{corollary}

\begin{proof}
Let $f\in \cC^\infty (B_\R)$ be arbitrary. Using (\ref{e1.3}) and the
$G_\R$-invariance of $\cB_\nu$ and $\pi$, we obtain
\begin{gather*}
    \int_{B_\R} dx\, h_\C (x,x)^{-p/2}  f(g x) (\ps_\nu F)(gx)
  = \int_{B_\R} dx\, h_\C (x,x)^{-p/2}  \ps_\nu ((f\circ \pi) F)(gx) \\
 \qquad{} = \int_{B_\R} dx\, h_\C (x,x)^{-p/2}  \ps_\nu ((f\circ \pi) F)(x)
  = \int_{B_\C} dz\, h_\C (z,z)^{-p}  \cB_\nu (z)
    (f\circ \pi)(z)  F(z) \\
\qquad{} = \int_{B_\C} dz\, h_\C (z,z)^{-p}  \cB_\nu (gz) (f\circ \pi)(gz)
    F(gz) \\
\qquad{} = \int_{B_\C} dz\, h_\C (z,z)^{-p}  \cB_\nu (z) (f\circ g)
    (\pi (z)) (F\circ g)(z) \\
\qquad{} = \int_{B_\R} dx\, h_\C (x,x)^{-p/2} \ps_\nu
    (((f\circ g)\circ \pi)(F\circ g))(x) \\
\qquad{} = \int_{B_\R} dx\,  h_\C (x,x)^{-p/2} (f\circ g)(x)  \ps_\nu
    (F\circ g)(x).\tag*{\qed}
\end{gather*}\renewcommand{\qed}{}
\end{proof}

For $x=0\in B_\R$ we have in particular
\begin{equation}\label{e1.4}
 (\ps_\nu F)(0) = \int_{Y\cap B_\C} dy \, |\det\Phi' (0,y)|
  \cB_\nu(y)  F(y)  h_\C (y,y)^{-p}.
\end{equation}
Our next goal is to obtain an asymptotic expansion of (\ref{e1.4}),
as~$\nu\to\infty$, using the method of stationary phase but also the
more ref\/ined ``$K_\R$-invariant'' Taylor expansion of $F$ at $0\in Y$.
As a~f\/irst step we recall that
\[
 Y = \Lambda  Z_\R = \{\Lambda x :\  x \in Z_\R\},
\]
for an $\R$-linear (but not necessarily $\C$-linear) isomorphism $\Lambda:
Z_\C\to Z_\C$ which commutes with~$K_\R$. For $f\in\cC^\infty(B_\R)$ we have
$f\circ \Lambda^{-1}\in \cC^\infty (Y \cap B_\C)$. Consider the distribution
\begin{equation}\label{e1.5}
  f \mapsto \ps_\nu \big(f\circ \Lambda^{-1}\big)(0)
\end{equation}
on $B_\R$, which by construction is $K_\R$-invariant. For any partition
$\bm\in\N^r_+$ let $E^\bm_\R$ be the $K_\R$-invariant
constant coef\/f\/icient dif\/ferential operator on $Z_\R$ corresponding to the
polynomial $E^\bm$ introduced in Section~\ref{SecPRELIM}. Using
multi-indices $\varkappa\in \N^d$ we may write
\[
 E^\bm(x) = \sum_\varkappa c^\bm_\varkappa   x^\varkappa,   \qquad
 E^\bm_\R = \sum\limits_\varkappa
 c^\bm_\varkappa  \partial^\varkappa_\R,
\]
where $\partial^\varkappa_\R$ is the ``real'' partial derivative operator
on $Z_\R$ associated with $\varkappa$ and $|\varkappa| \leq |\bm|$.
Expressing $\partial^\varkappa_\R$ in terms of Wirtinger type derivatives
$\partial^\sigma_\C$, $\overline \partial^{\tau}_\C$ on $Z_\C$, for
multi-indices \mbox{$\sigma, \tau \in \N^d$}, such that $|\sigma|\leq |\bm|
\geq |\tau|$, $E^\bm_\R$~determines a complexif\/ied constant coef\/f\/icient
dif\/ferential operator
\[
 E^\bm_\C = \sum\limits_{\sigma, \tau}
 c^\bm_{\sigma, \tau}  \partial^\sigma_\C
 \overline{\partial}^{\tau}_\C
\]
for suitable constants $c^\bm_{\sigma, \tau} \in \C$.
Pulling back by the (real-linear) map $\Lambda$ we get
\[
 \partial^\sigma_\C \overline{\partial}^{\tau}_\C (F\circ \Lambda) =
 \sum\limits_{\alpha, \beta} \Lambda^{\sigma, \tau}_{\alpha, \beta}
  (\partial^\alpha_\C  \overline \partial{}^\beta_\C  F) \circ\Lambda
\]
for suitable constants $\Lambda^{\sigma, \tau}_{\alpha, \beta} \in \C$,
and hence
\begin{gather*}
 E^\bm_\C (F\circ \Lambda)
= \sum\limits_{\sigma,\tau} c_{\sigma, \tau}^\bm
     \partial^\sigma_\C   \overline{\partial}^{\tau}_\C (F\circ \Lambda)
 = \sum\limits_{\sigma, \tau}  c_{\sigma, \tau}^\bm
    \sum\limits_{\alpha, \beta} \Lambda^{\sigma,\tau}_{\alpha, \beta}
     (\partial^\alpha_\C   \overline{\partial}^{\beta}_\C   F) \circ\Lambda
= \sum\limits_{\alpha, \beta} P^\bm_{\alpha,\beta}
     (\partial^\alpha_\C  \overline \partial{}^{\beta}_\C   F) \circ\Lambda,
\end{gather*}
where
\begin{equation}\label{e1.6}
 P^\bm_{\alpha,\beta} =
 \sum\limits_{\sigma,\tau} c^\bm_{\sigma,\tau}
 \  \Lambda^{\sigma,\tau}_{\alpha,\beta}.
\end{equation}
Returning to the distribution (\ref{e1.5}) on $B_\R$, one has

\begin{proposition}\label{p1.1}
There exist unique constants $[\nu]_\bm$, for $\bm\in\N^r_+$,
such that for all $F\in \cC^\infty (B_\C)$
\begin{equation}\label{e1.7A}
 (\ps_\nu\, F)(0) \sim \sum\limits_\bm
   \frac{1}{[\nu]_\bm}
  E^\bm_\C  (F\circ \Lambda)(0)
\end{equation}
as an asymptotic expansion.
\end{proposition}

\begin{proof} By~the def\/inition of~$Y$, the~real-linear operator
$y\mapsto y^\#$ from $Y$ into $Z_\C$ is injective, and thus bounded below.
It~follows that also the ($G_\C$-invariant) pseudohyperbolic distance
\[ \rho(y,y^\#) := \|\gamma_y(y^\#)\| , \qquad y\in B_\C ,   \]
is~bounded below by a multiple of $\|y-y^\#\|$ if $y\in Y$. Since, by~the
familiar transformation rule for the Jordan determinant~$h_\C$,
\[ \cB_\nu(z)^2 = \frac {h(z,z)^\nu h(z^\#,z^\#)^\nu} {|h(z,z^\#)|^{2\nu}}
 = h(\gamma_z z^\#,\gamma_z z^\#)^\nu  \]
and $h(w,w)\le1$ on the closure of~$B_\C$, with equality if and only if $w=0$,
it~follows that $\cB_\nu$ has a global maximum on $Y$ at $y=0$, which also
dominates the boundary values of $\cB_\nu$ in the sense that $\cB_\nu(y_k)
\to1$, $y_k\in Y$, implies that $y_k\to0$. We~may therefore apply the method
of stationary phase exactly  as in Section~3 of \cite{EU} to
conclude that for any $F\in\cC^\infty(B_\C)$, for which the right-hand side
exists for some $\nu>p-1$, the integral
\begin{gather*}
   \ps_\nu F(0)  = \int_Y   F(y)   |\det\Phi'(0,y)|   \cB_\nu(y)
    h_\C(y,y)^{-p/2} \, dy  \\
\phantom{\ps_\nu F(0)}{} = |\det\Lambda| \int_{B_\R}   F(\Lambda x)   |\det\Phi'(0,\Lambda x)|
     \cB_\nu(\Lambda x)   h_\C(\Lambda x,\Lambda x)^{-p/2} \, dx
\end{gather*}
has an asymptotic expansion as $\nu\to+\infty$
\[ \ps_\nu F(0) \sim \nu^{-d/2} \sum_{k\ge0} S_k(\partial_\R)
 (F\circ\Lambda)(0)   \nu^{-k}  \]
for some constant coef\/f\/icient dif\/ferential operators~$S_k(\partial_\R)$,
with $S_k$ polynomials on~$Z_\R$. Since $\ps_\nu$ is $K_\R$-invariant,
so~must be the~$S_k$; thus they admit a decomposition
\[ S_k = \sum_{|\bm|\le k} q_{k\bm}   E^\bm, \qquad q_{k\bm}\in\C,  \]
into the ``even'' Peter--Weyl components~$E^\bm$. Interchanging the two
summations and setting
\[ 
\frac1{[\nu]_\bm} := \nu^{-d/2} \sum_k q_{k\bm}   \nu^{-k},  \]
the claim follows.
\end{proof}

Using the transvections $\gamma_x\in G_\R$, for $x\in B_\R$, we def\/ine a
$G_\R$-invariant dif\/ferential operator
\[
 \ps^\bm : \ \cC^\infty (B_\C) \to \cC^\infty (B_\R)
\]
by putting
\begin{equation}\label{e1.7B}
 \ps^\bm  (F)(x)
 := E^\bm _\C (F\circ \gamma_x \circ \Lambda)(0)
  = \sum\limits_{\alpha, \beta} P^\bm_{\alpha,\beta}
   \partial^\alpha_{\C}  \overline{\partial}^{\beta}_\C (F\circ \gamma_x)(0).
\end{equation}
Since $\gamma_x : B_\C\to B_\C$ is holomorphic, there exist smooth functions
$\gamma^\alpha_\iota :   B_\R\to \C$, with $|\iota| \leq |\alpha|$, such that
\[
 \partial^\alpha_\C (H \circ \gamma_x)(0) = \sum\limits_\iota
 \gamma^\alpha_\iota (x) (\partial^{\, \iota}_\C  H)(x)
\]
for all $H\in \cO (B_\C)$ and $x\in B_\R$. Since $\ps_\nu$ is $G_\R$-invariant,
Proposition \ref{p1.1} implies
\begin{equation}\label{e1.7C}
 (\ps_\nu\, F)(x) \sim \sum\limits_\bm
 \frac{1}{[\nu]_\bm}
   (\ps^\bm\, F)(x)
\end{equation}
for all $F\in \cC^\infty (B_\C)$ and $x\in B_\R$.

Now let $\bm\in \N^r_+$ and $\varkappa\in \N^d$ be f\/ixed, with $|\varkappa|
\leq |m|$. Def\/ine a (non-invariant) ``holomorphic'' dif\/ferential operator
\[
 \ps^\bm_\varkappa : \ \cO (B_\C)\to \cC^\infty (B_\R)
\]
by the formula
\begin{equation}\label{e1.8}
  (\ps^\bm_\varkappa\, H)(x)
 =  \sum\limits_{\alpha, \beta} P^\bm_{\alpha,\beta}
    \partial^\alpha_\C  (H\circ \gamma_x)(0)
   \overline{\gamma^\beta_\varkappa(x)}
 =  \sum\limits_{\alpha, \beta, \iota}
  P^\bm_{\alpha,\beta}  \gamma^\alpha_\iota (x)
  \overline{\gamma^\beta_\varkappa (x)}\ (\partial^{\, \iota}_\C  H)(x)
\end{equation}
for all $x\in B_\R$ and $H\in \cO (B_\C)$, where the constants
$P^\bm_{\alpha,\beta}$ are def\/ined by (\ref{e1.6}).

\begin{lemma}\label{l1.1}
Let $G,H\in \cO(B_\C)$. Then
\[
 \ps^\bm (\overline G H)(x)
 = \sum_\varkappa (\ps^\bm_\varkappa H)(x)
  \overline{\partial^\varkappa_\C  G (x)}.
\]
\end{lemma}

\begin{proof}
Since $\gamma_x$ preserves holomorphy, (\ref{e1.8}) implies
\begin{gather*}
   \ps^\bm (\overline G H)(x)
   = \sum\limits_{\alpha, \beta} P^\bm_{\alpha, \beta}
       \partial^\alpha_\C   \overline{\partial}_\C^{\beta}
     (\overline{G\circ \gamma_x}  (H\circ \gamma_x)) (0) \\
  = \sum_{\alpha,\beta}  P^\bm_{\alpha,\beta}  \partial^\alpha_\C
     (H\circ\gamma_x)(0)  \overline{\partial^\beta_\C (G\circ \gamma_x)(0)}
   = \sum_{\alpha,\beta,\iota,\varkappa}
     P^\bm_{\alpha,\beta}  \gamma^\alpha_\iota (x)
     (\partial^{\, \iota}_\C H)(x)  \overline{\gamma^\beta_\varkappa (x)
     (\partial^\varkappa_\C G)(x)} \\
   = \sum_{\alpha,\beta,\iota,\varkappa}
     P^\bm_{\alpha,\beta}  \gamma^\alpha_\iota (x)
       \overline{\gamma^\beta_\varkappa (x)}  (\partial^{\, \iota}_\C H)(x)
       \overline{(\partial^{\varkappa}_\C G)(x)}
   = \sum_\varkappa (\ps^\bm_\varkappa  H)(x)
      \overline{\partial^{\varkappa}_\C G(x)}.\tag*{\qed}
\end{gather*}\renewcommand{\qed}{}
\end{proof}

\begin{definition}\label{d2.9}
For $\bm \in \N^r_+$, the $\bm$-th \textit{Moyal component} is the dif\/ferential
operator
\[
 \rho^\bm : \ \cO (B_\C) \to \cC^\infty (B_\R)
\]
def\/ined by the formula
\begin{equation}\label{e1.9A}
 (\rho^\bm\, H) (x) = h_\C(x,x)^{p/2}
 \sum\limits_\varkappa (-1)^\varkappa  \partial^\varkappa_\R (h_\C^{-p/2}
  \ps^\bm_\varkappa  H)(x)
\end{equation}
for all $x\in B_\R$ and $H\in \cO (B_\C)$. Here $\partial^\varkappa_\R$
is the ``real'' partial derivative operator on $B_\R\subset Z_\R$, and
$(-1)^\varkappa\, \partial^\varkappa_\R$ is its (Euclidean) adjoint.
We~also write just $h_\C$ for $h_\C(x,x)$.
\end{definition}

\begin{proposition}\label{p1.10}
Let $G, H \in \cO(B_\C)$. Then
\begin{equation}\label{e1.9B}
 \int_{B_\R} dx \, h_\C(x,x)^{\frac{\nu-p}{2}}  \overline{G(x)}
 (\rho^\bm\, H)(x)
 = \int_{B_\R} dx \, h_\C(x,x)^{-p/2}  \ps^\bm
 (\overline{G/I_\nu}  H)(x).
\end{equation}
\end{proposition}

\begin{proof}
Since $G$ is holomorphic, we have $\partial^\varkappa_\C\, G (x) =
\partial^\varkappa_\R\, G (x)$ for all $\varkappa \in \N^d$ and $x\in B_\R$.
Applying Lemma \ref{l1.1} to $G/I_\nu$ we obtain
\begin{gather*}
   \int_{B_\R} dx\, h_\C(x,x)^{\frac{\nu -p}{2}}  \overline{G(x)}
      (\rho^\bm H)(x)
  = \int_{B_\R} dx\, h_\C (x,x)^{-p/2}  \overline{(G/I_\nu) (x)}
      (\rho^\bm H)(x)\\
  = \sum\limits_{\varkappa}  \int_{B_\R} dx\,  \overline{(G/I_\nu) (x)}
    (-1)^\varkappa\ \partial^\varkappa_\R  (h_\C^{-p/2}
    \ps^\bm_\varkappa H)(x) \\
  = \sum\limits_{\varkappa}  \int_{B_\R} dx\,
    \overline{\partial^\varkappa_\R (G/I_\nu)(x)}  h_\C (x,x)^{-p/2}
     (\ps^\bm_\varkappa H)(x) \\
  = \sum\limits_{\varkappa}  \int_{B_\R} \! dx\,
    \overline{\partial^\varkappa_\C (G/I_\nu)(x)}  h_\C(x,x)^{-p/2}
      (\ps^\bm_\varkappa H)(x)
 =  \int_{B_\R} \! dx\, h_\C (x,x)^{-p/2}
    \ps^\bm (\overline{G/I_\nu}  H)(x) .\!\!\!\!\!\!\!\!\tag*{\qed}
\end{gather*}\renewcommand{\qed}{}
\end{proof}

As a consequence of Proposition \ref{p1.10} we obtain

\begin{corollary}
The~differential operators $\rho^\bm$ are
$G_\R$-invariant, i.e.,
\[
 \rho^\bm   (H\circ g)(x)
 = (\rho^\bm  H)(g(x))
\]
for all $H\in \cO(B_\C)$, $ g\in G_\R$ and $x\in B_\R$.
\end{corollary}

\begin{proof}
Replacing $G/I_\nu = \Phi$, (\ref{e1.9B}) can be written as
\[
 \int_{B_\R} dx\, h_\C(x,x)^{-p/2}  \overline{\Phi(x)}
  (\rho^\bm  H)(x)
 = \int_{B_\R} dx\, h_\C(x,x)^{-p/2}
    \ps^\bm  (\overline \Phi H)(x)
\]
for $\Phi, H \: {\in}\: \cO(B_\C)$. Since $\ps^\bm$ is $G_\R$-invariant
by construction (cf.~(\ref{e1.7B})), the assertion   follows. \end{proof}

The~main result of this section yields the desired asymptotic expansion
of the Moyal type restriction operator $\rho_\nu$ in terms of the invariant
dif\/ferential operators $\rho^\bm$:

\begin{theorem}\label{t2.12}
For $H\in \cO(B_\C)$ we have an asymptotic expansion
\[
 \rho_\nu H \sim \sum\limits_\bm
   \frac{1}{[\nu]_\bm}
   \rho^\bm H,
\]
where $\rho^\bm :   \cO (B_\C) \to \cC^\infty (D_\R)$
are $G_\R$-invariant holomorphic differential operators independent of
$\nu$ and the constants $[\nu]_\bm$ are determined by~\eqref{e1.7A}.
\end{theorem}

\begin{proof}
Let $G, H\in \cO (B_\C)$. Applying Proposition \ref{p1.2} and \ref{p1.5},
we obtain with (\ref{e1.7C}) and (\ref{e1.9B})
\begin{gather*}
   \int_{B_\R} dx\, h_\C(x,x)^{\frac{\nu-p}{2}}  \overline{G(x)}
      (\rho_\nu H)(x)
  = \int_{B_\C} dz\, h_\C(z,z)^{-p}   \cB_\nu (z)
      \overline{(G/I_\nu) (z)}  H(z) \\
 = \int_{B_\R} dx\, h_\C(x,x)^{-p/2}  \ps_\nu
      (\overline{G/I_\nu}   H)(x)
  = \sum\limits_\bm
      \frac{1}{[\nu]_\bm}
      \int_{B_\R} dx\, h_\C (x,x)^{-p/2}
      \ps^\bm\, (\overline{G/I_\nu}   H) (x) \\
  = \sum\limits_\bm
     \frac{1}{[\nu]_\bm}
     \int_{B_\R} dx\, h_\C (x,x)^{\frac{\nu-p}{2}}
      \overline{G(x)}  (\rho^\bm H)(x).
\end{gather*}
Since $G\in \cO(B_\C)$ is arbitrary, the assertion follows.
\end{proof}

\begin{remark} \label{RemGe}
Most --  probably all  -- of~the above extends also to the case of general
$G_\R$-invariant smooth retractions $\pi: B_\C\to B_\R$, i.e.~when $\pi^{-1}
(0)$ is not necessarily an intersection of~$B_\C$ with some real subspace~$Y$,
or~that the parameterization $\Lambda:B_\R\to\pi^{-1}(0)$ is not necessarily
linear but only smooth. In~fact, the application of the stationary
phase method in the proof of Proposition~\ref{p1.1} involves only the germs
of $F$ and~$\pi^{-1}(0)$ (or,~equivalently,~$\Lambda$) at~the origin. Thus
 we may replace the variety $\pi^{-1}(0)$ by its tangent space
at $0\in\pi^{-1}(0)$, and $\Lambda:B_\R\to\pi^{-1}(0)$ by~its dif\/ferential
at the origin. We~omit the details.
\end{remark}

\section{Asymptotic expansion: the complex case}
\label{SecDOUBLE}

In~the complex  case, where
\begin{gather*}
 B_\C   =   B  \times  \overline B = \{(z,\overline w) : \ z,w \in B\}, \\
 B_\R   =   \{(z,\overline z) : \ z \in B\}
\end{gather*}
and $B$ is an irreducible complex Hermitian bounded symmetric domain
(of rank $r$), the Moyal type restriction operator
\[
 \rho_\nu : \ \cC^\infty (B_\C) = \cC^\infty (B) \, \overline{\otimes} \,
   \cC^\infty(\overline B) \to \cC^\infty (B_\R)
\]
can be identif\/ied with the Moyal type (star-) product $\sharp_\nu$ via the
formula
\[
 \rho_\nu (f\otimes g) = f\, \sharp_\nu\,  g
\]
for all $f,g \in \cC^\infty (B)$. In~this case an asymptotic expansion has
been constructed in~\cite{E}, and here we show that the general construction
described in Section~\ref{SecMOYAL} yields precisely the expansion of~\cite{E}.
This is not completely obvious, since the construction in \cite{E} is based
on the complex structure of $B$ whereas the general construction of
Section~\ref{SecMOYAL} uses the ``real'' structure of $B_\R$.

The~f\/irst step is to identify the Berezin kernel $\cB_\nu$ on $B_\C$,
def\/ined in (\ref{e1.1}), for the complex  case. We have
\[
 h_\C ((z,\overline w), (\zeta,\overline \omega)) = h(z,\zeta)  h(\omega, w)
\]
for $z,w,\zeta,\omega\in B$ and the involution is given by
\[
 (z, \overline w)^\sharp := (w, \overline z).
\]
Therefore
\[
 \cB_\nu (z, \overline w)
 =  \frac{h_\C ((z, \overline w), (z, \overline w))^\nu}
         {|h_\C ((z, \overline w), (w, \overline z))|^\nu}
 =  \frac{h (z,  z)^\nu   h (w,  w)^\nu}{h (z,  w)^\nu\ h (w,  z)^\nu}
\]
coincides with the integral kernel for the $G$-invariant
{\it Berezin transform}
\[
 \cB_\nu : \ \cC^\infty (B) \to \cC^\infty (B)
\]
on $B$. This is clearly invariant under
\[
 G_\R = \{(g,g) : \ g\in G\},
\]
where $G = \Aut\, (B)$.
The~construction in \cite{E} starts with the asymptotic expansion
\[
 (\cB_\nu   f)(0) = \int_B dz\, h(z,z)^{\nu-p}  f(z)
 = \sum\limits_\bm
     \frac{1}{(\nu)_\bm}
     (E^\bm_\R   f)(0)
\]
of the $\nu$-Berezin transform $\cB_\nu$ associated with the usual
Toeplitz--Berezin quantization of $B$. Here, for any partition $\bm\in\N^r_+$,
the Pochhammer symbol
\[
 (\nu)_\bm := \frac{\Gamma_\Omega
 (\nu+ \bm)}{\Gamma_\Omega (\nu)}
 \]
is def\/ined via the Koecher--Gindikin $\Gamma$-function, and the
``sesqui-holomorphic'' constant coef\/f\/i\-cient dif\/ferential operators
$E^\bm_\R$ are def\/ined via the Fock space expansion
\[
 e^{(z|w)} = \sum\limits_\bm   E^\bm (z,w)
\]
for all $z,w\in Z$. In~multi-index notation,
\begin{equation}\label{e2.1}
 E^\bm(z,w) = \sum_{\alpha,\beta}   c^\bm_{\alpha,\beta}
    z^\alpha   \overline w{}^\beta,  \qquad
 E^\bm_\R = \sum\limits_{\alpha, \beta}     c_{\alpha\,\beta}^\bm
   \partial^\alpha\, \overline \partial{}^\beta
\end{equation}
for suitable constants $c_{\alpha\, \beta}^\bm$ and multi-indices
$\alpha,\beta\in \N^d$, such that $|\alpha| \leq |\bm| \geq |\beta|$. Since
\[
 \overline{E^\bm (z,w)}
 = E^\bm (w,z)
\]
it follows that
\begin{equation}\label{e2.2}
 \overline{ c_{\alpha\, \beta}^\bm}
 =  c_{\beta\, \alpha}^\bm.
\end{equation}
Passing to the complexif\/ication $Z_\C = Z\times \overline Z$, with variables
$(z,\overline w)$ for $z,w\in Z$, we use pairs of multi-indices and write
$\partial_\C^{\alpha \overline \beta}$ and $\overline \partial_\C^{\gamma
\overline \delta}$ for the associated Wirtinger derivatives. Thus, for
functions on $B_\C$ of the form
\begin{equation}\label{e2.2a}
  (f\otimes g)(z,\overline w) = f(z) g(w),
\end{equation}
we have
\begin{equation}\label{e2.3}
   \partial^{\alpha \overline \beta}_\C
    \overline \partial^{\gamma \overline \delta}_\C  (f\otimes g)
 = (\partial^{\alpha}  \overline \partial{}^\gamma f)
   \otimes (\overline \partial{}^\beta  \partial^\delta  g).
\end{equation}
Note that the f\/irst and second variable are treated dif\/ferently, since
holomorphic functions on~$B_\C$ correspond to the case where $f$ is
holomorphic and $g$ is anti-holomorphic. Let
\[
 \Lambda : \ B_\R \to B_\C
\]
denote the $\R$-linear mapping
\[
 \Lambda (z, \overline z) = (z,0)
\]
which is clearly $K_\R$-invariant. Consider the $G_\R$-invariant retraction
\[
 \pi : \ B_\C \to B_\R
\]
def\/ined by $\pi (z,\overline w) := (w,\overline w)$. Then
\[
\pi^{-1} (0) = \{(z,0) : \ z \in B\} = \Lambda\, B_\R.
\]

\begin{lemma}
For $F\in \cC^\infty (B_\C)$ we have
\[
 E^\bm_\C  (F \circ \Lambda)(0)
 = \sum\limits_{\alpha,\beta}
     c^\bm_{\alpha\, \beta}
    \partial^{\alpha 0}_\C    \overline \partial{}^{\beta 0}_\C  F (0).
\]
\end{lemma}

\begin{proof}
We may assume that $F(z,\overline w) = f(z)  g(w)$ is of the form
(\ref{e2.2a}) . Since
\[
 ((f\otimes g)\circ \Lambda)(z, \overline z) = (f\otimes g)(z,0) = f(z)  g(0)
\]
it follows from (\ref{e2.3}) and (\ref{e2.1}) that
\begin{gather*}
   E^\bm_\C   ((f\otimes g)\circ \Lambda) (0)
 = (E^\bm  f)(0)  g(0)
 = \sum\limits_{\alpha,\beta}
      c^\bm_{\alpha\, \beta}
   (\partial^\alpha  \overline \partial{}^\beta  f)(0)  g(0)\\
\phantom{E^\bm_\C   ((f\otimes g)\circ \Lambda) (0)}{}
 = \sum\limits_{\alpha,\beta}
     c^\bm_{\alpha\, \beta}
    (\partial^{\alpha\, 0}_\C  \overline \partial{}^{\beta\, 0}_\C
     (f\otimes g))(0).\tag*{\qed}
\end{gather*}\renewcommand{\qed}{}
\end{proof}

Comparing with the coef\/f\/icients $P^\bm_{\alpha,\beta}$ introduced
by (\ref{e1.6}) in the general case, it follows that
\begin{equation}\label{e2.4}
  P^\bm_{\alpha 0, \beta 0}
 = c^\bm_{\alpha\, \beta}
\end{equation}
for $\alpha,\beta\in \N^d$, whereas all other such coef\/f\/icients vanish.
This ref\/lects the fact that $\Lambda$ is trivial on the second component.
For~$z\in B$, let as before $\gamma_z \in G$ be the transvection mapping
$0$ to $z$. Then we have for $\alpha \in \N^d$ and $f\in \cO(B)$
\[
 \partial^\alpha (f\circ\gamma_z)(0) = \sum\limits_{\iota \leq \alpha}
 \gamma^\alpha_\iota (z)(\partial^{\, \iota}  f)(z),
\]
where $\gamma^\alpha_\iota$ are smooth functions on $B$. As~in \cite{E}
def\/ine a $G$-invariant operator
\[
 \cE^\bm : \ \cC^\infty  (B) \to \cC^\infty (B)
\]
by putting
\[
  (\cE^\bm\, f)(z) := E^\bm_\R (f\circ \gamma_z)(0).
\]
Then we have for $f,g\in \cO(B)$
\begin{gather}
   (\cE^\bm  (f \overline g))(z)
  = E^\bm_\R ((f\overline g)\circ \gamma_z)(0)
  = E^\bm_\R ((f\circ \gamma_z)   \overline{g\circ \gamma_z})(0) \nonumber\\
\phantom{(\cE^\bm  (f \overline g))(z)}{} = \sum\limits_{\alpha,\beta}
      c^\bm_{\alpha\, \beta}
      \partial^\alpha  \overline \partial{}^\beta  ((f\circ \gamma_z)
     \overline{g\circ \gamma_z})(0)
  = \sum\limits_{\alpha,\beta}
      c^\bm_{\alpha\, \beta}
     \partial^\alpha (f\circ \gamma_z)(0)  \overline{\partial^\beta
     (g\circ \gamma_z)(0)} \nonumber\\
\phantom{(\cE^\bm  (f \overline g))(z)}{}
  = \sum\limits_{\alpha,\beta}  \sum\limits_{\varkappa, \iota}
      c^\bm_{\alpha\, \beta}
     \gamma^\alpha_\varkappa (z) (\partial^\varkappa  f)(z)
      \overline{\gamma^\beta_\iota (z)}
      \overline{(\partial^{\, \iota}  g)(z)}.
  \label{e2.5a}
\end{gather}
Following \cite[Section~4]{E} one def\/ines (non-invariant) dif\/ferential
operators $\cR^\bm_\varkappa$, for any partition $\bm\in\N^r_+$ and any
multi-index $\varkappa \in \N^d$ with $|\varkappa| \leq |\bm|$, via the
expansion
\[
 (\cE^\bm  (f\overline g))(z)
 = \sum\limits_\varkappa(\partial^\varkappa  f)(z)
   (\cR^\bm_\varkappa   \overline g)(z),
\]
where $f\in\cO(B)$, $g\in\cC^\infty(B)$.  Comparing with \eqref{e2.5a} it
follows that
\[
 (\cR^\bm_\varkappa  \overline g)(z)
 = \sum\limits_{\alpha,\beta}   \sum\limits_{\iota}
    c^\bm_{\alpha\, \beta}
    \gamma^\alpha_\varkappa (z)  \overline{\gamma^\beta_\iota (z)}
    \overline{(\partial^{\, \iota}  g)(z)}
\]
whenever $g$ is holomorphic. On~the other hand, putting
\[
 \gamma_{z,\overline z} := (\gamma_z, \gamma_z)\in G_\R\subset G \times G
\]
we have for the ``holomorphic'' Wirtinger derivatives
\begin{gather}
    \partial^{\alpha\, \overline \beta}_\C
      [(f\otimes \overline g)\circ \gamma_{z,\overline z}](0,0)
  = \partial^{\alpha\, \overline \beta}_\C
     [(f\circ \gamma_z)\otimes \overline{g\circ \gamma_z}](0,0)
  = \partial^\alpha (f\circ \gamma_z)(0)  \overline{\partial^\beta
    (g\circ \gamma_z)(0)} \nonumber \\
 \qquad{} = \sum\limits_{\varkappa,\iota}   \gamma^\alpha_\varkappa (z)
    (\partial^{\varkappa}  f)(z)  \overline{\gamma^\beta_\iota (z)}
    \ \overline{\partial^{\, \iota}  g (z)}
  = \sum\limits_{\varkappa,\iota}  \gamma^\alpha_\varkappa (z)
     \overline{\gamma_\iota^\beta (z)}
      \partial^{\varkappa\, \overline \iota}_\C
     (f\otimes \overline g)(z, \overline z).   \label{e:2.7}
\end{gather}
We will now compute the (non-invariant) operators $\ps^\bm_\varkappa$,
introduced in (\ref{e1.8}), for the complex case.
Combining (\ref{e2.4}) and (\ref{e:2.7}) it follows that the non-zero operators
correspond to multi-index pairs $(\varkappa, 0)$ for $\varkappa \in \N^d$ and,
in view of (\ref{e2.2}),
\begin{gather*}
   \ps^\bm_{\varkappa 0}
      (f \otimes \overline g)(z, \overline z)
  = \sum\limits_{\alpha,\beta,\iota}
      P^\bm_{\alpha 0,\, \beta 0}
      \gamma^\alpha_\iota (z)  \overline{\gamma^\beta_\varkappa (z)}
      \partial^{\iota 0}_\C  (f \otimes \overline g)(z, \overline z) \\
\phantom{\ps^\bm_{\varkappa 0}
      (f \otimes \overline g)(z, \overline z)}{} = \sum\limits_{\alpha,\beta,\iota}
     c^\bm_{\alpha\, \beta}
      \gamma^\alpha_\iota (z)  \overline{\gamma^\beta_\varkappa (z)}
     (\partial^{\, \iota}  f)(z)\ \overline{g (z)}
  = \overline{(\cR^\bm_\varkappa
      \overline f)(z)  g(z)}.
\end{gather*}
This passing to the complex conjugate (also in the proof of the following
Proposition) could be avoided by working  with the ``anti-holomorphic''
second variable instead.

The~$G$-invariant bi-dif\/ferential operators $A_\bm$
on~$B$, introduced in \cite[Section~4]{E}, satisfy
\[
 A_\bm (f, \overline g)(z)
 = \sum\limits_\varkappa h(z,z)^p (-\partial)^\varkappa
 (h^{-p}\, f(\cR^\bm_\varkappa
 \, \overline g))(z)
\]
for all $f, g\in \cO(B)$, and are uniquely determined by this property
since $A_\bm$ involves only holomorphic derivatives in the f\/irst variable
and anti-holomorphic derivatives in the second variable.
By~\cite[Proposition~6]{E},
\[
 \overline{A_\bm (f, \overline g)}
 = A_\bm (g, \overline f)
\]
for all $f,g\in \cO(B)$.

\begin{proposition}\label{p2.2}
Let $f,g\in \cO(B)$. Then
\[
 \rho^\bm  (f\otimes \overline g)(z,\overline z)
 = A_\bm  (f, \overline g)(z)
\]
for all $z\in B$.
\end{proposition}

\begin{proof}
Since the operators $\rho^\bm$ are def\/ined by taking
suitable adjoints on $B_\R$, which requires another  identif\/ication, we instead
verify that both operators satisfy the same integral duality formula. Thus let
$f, g, \phi, \psi\in \cO (B)$. Then, by~(\ref{e1.9A}),
\begin{gather*}
   \int_B dz\, h(z,z)^{-p}  \overline{\phi (z)}\ \psi(z)
       \rho^\bm
     (f\otimes \overline g)(z, \overline z) \\
   = \int_{B_\R} d(z,\overline z)
     \, h_\C ((z,\overline z), (z,\overline z))^{-p/2}
       \overline{(\phi\otimes \overline \psi)(z,\overline z)}
       \rho^\bm (f\otimes \overline g) (z,\overline z) \\
  = \sum\limits_\varkappa  \int_{B_\R} d(z,\overline z)
     \, h_\C ((z,\overline z), (z,\overline z))^{-p/2}
       \overline{\partial^{\varkappa 0}_\C  (\phi \otimes \overline \psi)
     (z,\overline z)}  \ps^\bm_{\varkappa 0}
     (f\otimes \overline g)(z,\overline z) \\
   = \sum\limits_\varkappa  \int_{B} dz\, h (z, z)^{-p}
      \overline{(\partial^\varkappa  \phi)(z)}\ \psi(z)
       \overline{g(z)  (\cR^\bm_{\varkappa}
       \overline f)(z)}   \\
  =  \sum\limits_\varkappa  \overline{\int_{B} dz\, h (z, z)^{-p}
        (\partial^\varkappa  \phi )(z)  \overline{\psi (z)}
       g(z) (\cR^\bm_\varkappa  \overline f)(z)} \\
  = \sum\limits_\varkappa  \overline{\int_{B} dz\, \phi (z)
       (-\partial)^\varkappa (h^{-p}  \overline \psi
       g (\cR^\bm_\varkappa  \overline f))(z)}
   = \sum\limits_\varkappa  \overline{\int_{B} dz\, \phi (z)
       \overline{\psi (z)}  (-\partial)^\varkappa  (h^{-p}
       g (\cR^\bm_\varkappa  \overline f))(z)} \\
   = \overline{\int_{B} dz\, h(z,z)^{-p}  \phi(z)  \overline{\psi(z)}
       A_\bm  (g,\overline f)(z)}
   = \int_{B} dz\, h(z,z)^{-p}  \overline{\phi(z)}\, \psi (z)
      A_\bm  (f,\overline g)(z).
    \end{gather*}
Since $\phi, \psi\in \cO(B)$ are arbitrary, the assertion follows.
Note that the formula (\ref{e1.9A}) def\/ining~$\rho^\bm$ uses the real
derivatives~$\partial_\R$, whereas in this section we are using rather
the Wirtinger derivatives~$\partial$ and~$\dbar$ on~$B$
(corresponding to viewing $B_\R=B$ as a domain in $\C^d$ rather
than~$\R^{2d}$); this is ref\/lected by the appearance of the Hermitian
adjoint $-\dbar{}^{\varkappa0}$ (rather than $-\partial^
{\varkappa0}$) of~$\partial^{\varkappa0}$ on the third line in the
computation above.
\end{proof}

\section{Examples}

\label{SecEXMPS}
We~begin with the case of the Euclidean space where everything can be computed
explicitly.

\begin{example} \label{ex1}
Let $B_\R=\R$, so~that $B_\C=\C$, and $\Lambda x:=\epsilon x$ for some
$\epsilon\in\C\setminus\R$, $|\epsilon|=1$. The~corresponding retraction
$\pi$ is just the oblique $\R$-linear projection associated to the direct
sum decomposition
\[ \C = \R \oplus \epsilon\R ;   \]
the mapping $\phi$ is just $\Phi(x,y)=x+y$, and $\det\Phi'=1$. The~role
of the Jordan determinant polynomial $h_\C(x,y)$ is played by the function
\be
 e^{-x\overline y}, \qquad x,y\in\C,  \label{e:TE}
\ee
and the ``genus'' $p=0$ while ``rank'' $r=1$.
The~partitions are just nonnegative integers $\bm=(m)$, and the polynomials
$E^\bm$ are simply
\[ E^\bm(x) = \frac{x^{2m}}{(2m)!}.  \]
Thus $E^\bm_\C=(\partial+\dbar)^{2m}/(2m)!$, and
\be \label{e:TQ}
 P^\bm_{\alpha,\beta} = \begin{cases}
 \dfrac{\epsilon^{\alpha-\beta}}{\alpha!\beta!}
   \quad &\text{if } \alpha+\beta=2m, \\  0 & \text{otherwise.}
 \end{cases}
\ee
The ``transvections'' $\gamma_x$ are just the ordinary translations
$\gamma_x y=x+y$, which implies that the functions $\gamma^\alpha_\iota$
equal constant one if $\alpha=\iota$, and vanish otherwise. Feeding all
this information into~(\ref{e1.8}) and~(\ref{e1.9A}), we~get
\[ \ps^\bm_\kappa = \frac{\epsilon^{2m-2\kappa}}{(2m-\kappa)!\kappa!}
 \; \partial^{2m-\kappa}  \]
and, for $H\in\cO(\C)$,
\[ \rho^\bm H = \frac1{(2m)!} \sum_{\kappa=0}^{2m} \binom{2m}\kappa
\epsilon^{2m-2\kappa} (-1)^\kappa \partial_\R^\kappa (\partial^{2m-\kappa}H)
= \frac{(\epsilon-\overline\epsilon)^{2m}}{(2m)!}   \partial^{2m}H.   \]
We~next compute the ``Pochhammer'' symbols~$[\nu]_\bm$, using the
formula~(\ref{e1.4}). By~(\ref{e:TE}),
\[
\ps_\nu F(0)
= \int_{\epsilon\R} F(y)   \frac{e^{-\nu|y|^2}}{|e^{-y^2}|^\nu} \, dy
= \int_\R F(\epsilon y)   e^{-\nu y^2 (1-\Re\epsilon^2)} \, dy.
\]
Denoting for brevity $1-\Re\epsilon^2=-\frac12(\epsilon-\overline\epsilon)^2
=:c>0$ and making the change of variable $y=x/\sqrt{c\nu}$ yields
\[ \ps_\nu F(0) = \frac1{\sqrt{c\nu}} \int_\R
F\left( \frac\epsilon{\sqrt{c\nu}}  x \right)   e^{-x^2} \, dx.   \]
We~may assume that $F$ is holomorphic; replacing then $F$ by its Taylor
expansion, integrating term by~term (which is easily justif\/ied), and using
the fact that $\int_\R x^{2j} e^{-x^2} \,dx=\Gamma(j+\frac12)$, we~f\/inally
arrive~at
\[ \ps_\nu F(0) = \frac1{\sqrt{c\nu}} \sum_{j=0}^\infty
 \left(\frac\epsilon{\sqrt{c\nu}}\right)^{2j}
 \frac{F^{(2j)}(0)}{(2j)!}   \Gamma(j+\tfrac12).   \]
As~$E^\bm_\C(F\circ\Lambda)(0)=\frac{\epsilon^{2j}}{(2j)!} F^{(2j)}(0)$,
we~thus~get
\be
 \frac1{[\nu]}_\bm = \frac{\Gamma(m+\frac12)}{(c\nu)^{m+\frac12}}
 = \frac{(2m)! \Gamma(\frac12)}{m!4^m(c\nu)^{m+\frac12}}   \label{e:TN}
\ee
where the last equality used the doubling formula for the Gamma function.

This corresponds to the unnormalized Lebesgue measure on~$\C$; it~is
usual to make a normalization so that $\rho_\nu{\mathbf1}={\mathbf1}$,
i.e.~$[\nu]_{(0)}=1$. If~this is done then (\ref{e:TN}) gets divided by
the same thing with $m=0$, that~is, it~becomes,
\[ \frac1{[\nu]_\bm} = \frac{\Gamma(m+\frac12)} {\Gamma(\frac12)(c\nu)^m}
= \frac{(2m)!}{(\epsilon-\overline\epsilon)^{2m} \nu^m m! (-2)^m}.  \]
Note that even though both $\rho^\bm$ and $[\nu]_\bm$ depend on~$\epsilon$,
the~sum
\[ \rho_\nu = \sum_\bm \frac{\rho^\bm}{[\nu]_\bm} = \sum_{m=0}^\infty
 \frac{\partial^{2m}} {m!(-2)^m\nu^m} = e^{-\partial^2/2\nu}  \]
is~independent of~it, as~it should.

A~similar analysis can be done for $B_\R=\R^d$, $d>1$; cf.~the next example.
\end{example}

\begin{example} \label{ex2}
$B_\R=\C^d\cong\R^{2d}$, so~that $B_\C=\C^d\times\overline{\C^d}$, where
as always we identify $B_\R$ with $\{(z,\oz): z\in\C^d\}\subset B_\C$.
For~$\Lambda$, we~let
\begin{gather}
 \Lambda z = (z,\overline{az})   \label{e:TL}
\end{gather}
with some f\/ixed $a\in\C$, $a\neq1$. The~retraction $\pi$ is the oblique
real-linear projection associated to the direct sum decomposition
\[ \C^d \times \overline{\C^d} = \Lambda\C^d \oplus \Lambda_1\C^d,   \]
where $\Lambda_1$ is as in (\ref{e:TL}) but with $a=1$. Using again the
Wirtinger derivatives $\partial$, $\dbar$ rather than~$\partial_\R$
on $\C^d\cong\R^{2d}$, we~have for any partition $\bm=(m)$
\[ E^\bm_\R = \sum_{|\beta|=m}
\frac{\partial^\beta \dbar{}^\beta} {\beta!}   \]
with the usual multi-index notation. Recalling that the numbers
$P^\bm_{\rho,\sigma}$ are, quite generally, def\/ined~by
\begin{gather}   \label{e:TP}
 \sum_{\rho,\sigma} P^\bm_{\rho,\sigma}   \partial^\rho_\C
 \dbar^\sigma_\C   F(0) = E^\bm_\R(F\circ\Lambda)(0),
\end{gather}
it~follows that
\be   \label{e:TS}
 P^\bm_{\alpha\overline\beta,\gamma\overline\delta}
 = \begin{cases} \dfrac{\rho!}{\alpha!\beta!\gamma!\delta!}
 a^{|\delta|} \overline a{}^{|\beta|} \quad & \text{if }
 \alpha+\delta = \beta+\gamma = \rho, \  |\rho|=m,  \\
 0 & \text{otherwise}.  \end{cases}
\ee
(Here we are again using the ``double'' Wirtinger derivatives
$\partial^{\alpha\overline\beta}_\C$ etc.~as in Section~\ref{SecDOUBLE}.)
As~in the preceding example, the role of the ``Jordan determinant'' $h_\C$
is played by the function
\be   \label{e:TH}
 h_\C((z,\ow),(z_1,\ow_1)) = e^{-(z|z_1)-(w_1|w)}
\ee
and $p=0$. Taking $H\in\cO(B_\C)$ of the form $H(z,\ow)=f(z)\overline{g(w)}$
with $f,g\in\cO(\C^d)$, we~get as in the preceding example
\begin{gather*}
\rho^\bm H  = \sum_{\kappa,\lambda} (-\dbar)^\kappa (-\partial)^\lambda
 \sum_{\alpha,\beta,\gamma,\delta} \sum_{\iota,\eta}
 P^\bm_{\alpha\overline\beta,\gamma\overline\delta}
 \gamma^{\alpha\overline\beta}_{\iota\overline\eta}
 \overline{\gamma^{\gamma\overline\delta}_{\kappa\overline\lambda}}
   \partial^{\iota\overline\eta} H  \\
\phantom{\rho^\bm H}{} = \sum_{\kappa,\lambda} (-\dbar)^\kappa (-\partial)^\lambda
 \sum_{\alpha,\beta}  P^\bm_{\alpha\overline\beta,\kappa\overline\lambda}
  \partial^{\alpha\overline\beta} H
= \sum_{\alpha,\beta,\kappa,\lambda} (-1)^{|\kappa+\lambda|}
 P^\bm_{\alpha\overline\beta,\kappa\overline\lambda}
 \partial^{\alpha+\lambda}f   \overline{\partial^{\beta+\kappa}g}  \\
\phantom{\rho^\bm H}{} = \sum_{|\rho|=m} \sum_{\beta,\lambda\le\rho} (-1)^{|\rho-\beta+\lambda|}
 \binom\rho\beta \binom\rho\lambda
   \frac{a^{|\lambda|} \overline a^{|\beta|}} {\rho!}
   \partial^\rho f   \overline{\partial^\rho g}  \\
\phantom{\rho^\bm H}{} = \sum_{|\rho|=m} \frac{(-1)^{|\rho|}}{\rho!}   |1-a|^{2|\rho|}
   \partial^\rho f   \overline{\partial^\rho g}
= \frac{(-1)^m}{m!}   |1-a|^{2m}
 \bigg( \sum_{j=1}^d \partial_{z_j} \dbar_{w_j} \bigg)^m   H .
\end{gather*}
(Here the appearance of $(-\dbar)^\kappa(-\partial)^\lambda$, rather than
$(-\partial)^\kappa(-\dbar)^\lambda$, is~for the same reason as indicated
at the end of the proof of Proposition~\ref{p2.2}.)  Thus, symbolically,
\[ \rho^\bm = \frac{(-1)^m}{m!}   |1-a|^{2m}   (\partial\otimes\dbar)^m. \]
To~compute $[\nu]_\bm$, we~again start from (\ref{e1.4}). Observe that for
the function $F\in\cO(B_\C)$ given~by
\[ F(z,\ow) = z^\alpha \oz^\beta w^\gamma \ow^\delta, \qquad \text{where} \quad
\alpha+\gamma=\beta+\delta=\rho,  \]
we~have by~(\ref{e:TP})
\[ E^\bm_\R (F\circ\Lambda)(0) = \alpha!\beta!\gamma!\delta!
P^\bm_{\alpha\overline\beta,\gamma\overline\delta}.  \]
Hence
\[ \ps_\nu F(0) = \sum_\bm   \frac1{[\nu]_\bm}   E^\bm_\C(F\circ\Lambda)(0)
= \frac{\rho!}{[\nu]_{|\rho|}}   a^{|\gamma|} \overline a^{|\delta|}.  \]
On~the other hand, from (\ref{e1.4}) and~(\ref{e:TH}),
\begin{gather*}
 \ps_\nu F(0)  = \int_{\C^d} F(z,\overline{az})   e^{-\nu|z|^2|1-a|^2} \,dz \\
 \phantom{\ps_\nu F(0)}{} = a^{|\gamma|} \overline a^{|\delta|} \int_{\C^d} z^\rho \oz{}^\rho
 e^{-\nu|1-a|^2|z|^2} \, dz
= \frac{\rho!}{(|1-a|^2\nu)^{|\rho|}}   a^{|\gamma|} \overline a^{|\delta|},
\end{gather*}
provided $dz$ is normalized so that $[\nu]_0=1$.
Thus (under this normalization)
\[ [\nu]_\bm = \nu^m   |1-a|^{2m}.   \]
Note that, again,
\[ \rho_\nu = \sum_\bm \frac{\rho^\bm}{[\nu]_\bm}
= \sum_{m=0}^\infty \frac{(-1)^m}{m!\nu^m}   (\partial\otimes\dbar)^m
= e^{-\partial\otimes\dbar/\nu}  \]
does not depend on~$a$, even though $\rho^\bm$ and $[\nu]_\bm$ both~do.
\end{example}

\begin{example}  \label{ex3}
As~a~f\/irst ``non-f\/lat'' situation, consider the unit interval $B_\R=(-1,+1)$
with complexif\/ication $B_\C=\D$, the unit disc in~$\C$; and we take the same
$\Lambda$ as in Example~\ref{ex1}, i.e.~$\Lambda x=\epsilon x$, $\epsilon\in
\mathbb T\setminus\R$. The constants $P^\bm_{\alpha,\beta}$ are thus still
given by~(\ref{e:TQ}), and $h_\C(x,y)=1-x\overline y$ while $p=2$. Thus for
$H\in\cO(\D)$,
\[ \rho^\bm H(x) = \big(1-x^2\big) \sum_\kappa (-1)^\kappa
 \Big(\frac d{dx}\Big)^\kappa \bigg(\frac1{1-x^2}
 \sum_{\begin{smallmatrix} \alpha,\beta,\iota \\
       \alpha+\beta=2m \end{smallmatrix}}
 \frac{\epsilon^{\alpha-\beta}}{\alpha!\beta!}
 \gamma^\alpha_\iota(x)   \overline{\gamma^\beta_\kappa(x)}
  \partial^\iota H(x) \bigg) .  \]
This time explicit formulas are hard to come~by, since the expressions
$\gamma^\alpha_\iota(x)$ are quite complicated. One~has, of~course,
$\rho^{(0)}H=H$, while
\[ \rho^{(1)}H(x) = (\epsilon-\overline\epsilon)^2
 \big[\big(1-x^2\big)^2 H''(x) - 2x\big(1-x^2\big) H'(x)\big] = (\epsilon-\overline\epsilon)^2
 (H\circ\gamma_x)''(0)  \]
is the $G_\R$-invariant operator uniquely determined by
$\rho^{(1)}H(0)=(\epsilon-\overline\epsilon)^2 H''(0)$.
Computer-aided calculation similarly gives
\begin{gather*}
\rho^{(2)}H(0)  = 24 (\epsilon-\overline\epsilon)^2 H''(0)
+ (\epsilon-\overline\epsilon)^4 H^{(4)}(0),   \\
\rho^{(3)}H(0)  = 1080 (\epsilon-\overline\epsilon)^2 H''(0)
+ 120 (\epsilon-\overline\epsilon)^4 H^{(4)}(0)
+ (\epsilon-\overline\epsilon)^6 H^{(6)}(0) .
\end{gather*}
The leading coef\/f\/icient in $\rho^{(m)}  H(0)$ is always $m^2 (2m-1)!$.

To~compute $[\nu]_\bm$, noting that $\det\Phi'(0,y)=1-y^2$ by
Corollary~\ref{coro1}, we~get from (\ref{e1.4}) and~(\ref{e1.7A}),
\[ \int_{-1}^1 F(\epsilon x)   |1-\epsilon^2 x^2|
 \bigg( \frac{1-x^2}{|1-\epsilon^2 x^2|} \bigg) ^\nu
 \frac{dx}{(1-x^2)^2} \sim \sum_\bm
 \frac{(\epsilon\partial+\overline\epsilon\dbar)^{2m}F(0)}
      {(2m)! [\nu]_\bm} .  \]
Denoting $F(\epsilon x)=:f(x)$ yields
\[ \int_{-1}^1 f(x)   \bigg( \frac{1-x^2}{|1-\epsilon^2x^2|} \bigg)^{\nu-1}
   \frac{dx}{1-x^2} \sim \sum_m   \frac{f^{(2m)}(0)}{(2m)![\nu]_\bm}.  \]
Taking in particular $f(x)=x^{2m}$ we obtain
\be   \label{e:TI}
 \frac1{[\nu]_\bm} = \int_{-1}^1 x^{2m}
   \frac{(1-x^2)^{\nu-2}} {|1-\epsilon^2 x^2|^{\nu-1}}  \, dx
 = \int_0^1 t^{m-\frac12}
  \frac{(1-t)^{\nu-2}}{|1-\epsilon^2 t|^{\nu-1}} \,dt.
\ee
Writing
\[ \frac1{|1-\epsilon^2 t|^{\nu-1}} = (1-\epsilon^2 t)^{-(\nu-1)/2}
 (1-\overline\epsilon{}^2 t)^{-(\nu-1)/2} = \sum_{j,k=0}^\infty  \frac
 {(\frac{\nu-1}2)_j (\frac{\nu-1}2)_k}{j!k!}   \epsilon^{2(j-k)} t^{j+k}  \]
we~arrive at the double series
\[ \frac1{[\nu]_\bm} =
 \frac{\Gamma(m+\frac12)\Gamma(\nu-1)} {\Gamma(m+\nu-\frac12)}
 \sum_{j,k\ge0} \frac{(\frac{\nu-1}2)_j (\frac{\nu-1}2)_k}{j!k!}
   \frac{(m+\frac12)_{j+k}}{(m+\nu-\frac12)_{j+k}}   \epsilon^{2(j-k)}.  \]
The~double sum on the right-hand side is the value at $x=\epsilon$,
$y=\overline\epsilon$ of the Horn hypergeometric function of two
variables \cite[\S~5.7.1]{BE}
\[ F_1\big(m+\tfrac12,\tfrac{\nu-1}2,\tfrac{\nu-1}2,m+\nu-\tfrac12,x,y\big)   \]
and in general cannot be evaluated in closed form. For~particular values
of~$\epsilon$, there may be some simplif\/ications; for~instance, for
$\epsilon=i$ the integral (\ref{e:TI}) becomes
\begin{gather*}  \frac1{[\nu]_\bm} = \int_0^1 t^{m-\frac12} (1-t)^{\nu-2} (1+t)^{1-\nu}
 \, dt \\
 \hphantom{\frac1{[\nu]_\bm}}{} = \frac{\Gamma(m+\frac12)\Gamma(\nu-1)}{\Gamma(m+\nu-\frac12)}
 \, {}_2\!F_1 \big(m+\tfrac12,\nu-1;m+\nu-\tfrac12;-1\big),
   \end{gather*}
where ${}_2\!F_1$ is the ordinary Gauss hypergeometric function.

We~remark that expressions involving values of ${}_2\!F_1$~at~$-1$ occur as
eigenvalues of the Berezin (or~``link'') transform corresponding to the Weyl
calculus on rank~1 real symmetric spaces, cf.~\cite[Theorem~4.1]{AUweyl}.
(Also, Horn's hypergeometric functions of another kind~-- namely, $\Phi_2$~in
the notation of~\cite{BE}~-- appear in the formula for the \emph{harmonic}
Segal--Bargmann kernel on~$\C^d$, see~\cite{E4}; it~is however unclear if there
is any deeper relationship.)
\end{example}

\begin{example}  \label{ex4}
In~this f\/inal example we consider $B_\R=\D$, embedded in $B_\C=\D\times
\overline\D$ in the usual way as $\{(z,\oz):z\in\D\}$. For~$\Lambda$ we
take the same map $\Lambda z=(z,\overline{az})$ as in Example~\ref{ex2},
with some f\/ixed $a\in\C$, $a\neq1$. The~corresponding retraction $\pi:
B_\C\to B_\R$ assigns to $(z,\ow)\in\D\times\overline\D$ the (unique)
point $x\in\D$ such that $\gamma_x w=a\gamma_x z$. (The~existence of
such $x$ follows by the following argument. For~any $z,w,u,v\in\D$,
the~existence of $g\in G$ such that $gz=u$, $gw=v$ is equivalent to
the equality
\be   \rho(z,w) = \rho(u,v)   \label{e:TR}   \ee
of the pseudohyperbolic distances $\rho(u,v):=|\frac{u-v}{1-\overline uv}|$.
On~the other hand, if~$u$ runs through the interval $[0,\min\{1,\frac1{|a|}\})$
and $v=au$, then $\rho(u,v)$ runs from $0$ to~$1$; thus (\ref{e:TR}) holds for
some~$u$. With $g$ as above, take $x=-g^{-1}(0)$.)    

The~constants $P^\bm_{\alpha\overline\beta,\gamma\overline\delta}$ are then
still given by the formula (\ref{e:TS}) from Example~\ref{ex2}, while the
corresponding functions $\gamma^{\alpha\overline\beta}_{\iota\overline\eta}$
are easily seen to be given~by $\gamma^{\alpha\overline\beta}_{\iota\overline
\eta}(z,\ow)=\gamma^\alpha_\iota(z) \overline{\gamma^\beta_\eta(w)}$, where
$\gamma^\alpha_\iota$ are the one-variable functions for the disc from the
preceding example. By~(\ref{e1.9A}) we thus get for $H(z,\ow)=f(z)\overline
{g(w)}$, $f,g\in\cO(\D)$, and $\bm=(m)$,
\begin{gather*}
 \rho^\bm H(z,\oz)  = (1-z\ow)^2 \sum_{\kappa,\lambda}
 (-\dbar_w)^\kappa (-\partial_z)^\lambda
 \bigg[ (1-z\ow)^{-2} \sum_{\alpha,\beta,\gamma,\delta,\iota,\eta}
 P^\bm_{\alpha\overline\beta,\gamma\overline\delta}   \\
 \phantom{\rho^\bm H(z,\oz)  =}{}  \hskip4em
 \gamma^\alpha_\iota(z) \overline{\gamma^\beta_\eta(w)}
 \overline{\gamma^\gamma_\kappa(w)} \gamma^\delta_\lambda(z)
   \partial^\iota f(z)   \overline{\partial^\eta g(w)} \bigg]   \Big|_{w=z}.
\end{gather*}
Here again $(-\dbar_w)^\kappa(-\partial_z)^\lambda$ occurs rather than
$(-\dbar_w)^\lambda(-\partial_z)^\kappa$, and likewise $\overline{\gamma
^\gamma_\kappa(w)}\gamma^\delta_\lambda(z)$ rather than $\gamma^\gamma
_\kappa(z)\overline{\gamma^\delta_\lambda(w)}$, for~the same reasons as
in Example~\ref{ex2} and in the proof of Proposition~\ref{p2.2}.

For~low values of~$m$, one~computes that $\rho^{(0)}(f\overline g)
=f\overline g$ (of~course), while
\[ \rho^{(1)}(f\overline g)(z) = -|1-a|^2  \big(1-|z|^2\big)^2
  f'(z)   \overline{g'(z)}  \]
is~the $G$-invariant operator from $\cO(\D\times\overline\D)$ into
$\cC^\infty(\D)$ uniquely determined~by
\[ \rho^{(1)}(f\overline g)(0) = -|1-a|^2   f'(0)  \overline{g'(0)}.  \]
Computer-aided calculations give
\[ \rho^{(2)}(f\overline g)(0) = -\frac{|1-a|^2}2   \Big[ 4(1+|a|^2)
 f'(0) \overline{g'(0)} - |1-a|^2 f''(0) \overline{g''(0)} \Big]  \]
and
\begin{gather*}
 \rho^{(3)}(f\overline g)(0)
  = -\frac{|1-a|^2}6 \, \Big[ 36(1+|a|^2+|a|^4) f'(0) \overline{g'(0)}
    - 18 |1-a|^2 (1+|a|^2) f''(0) \overline{g''(0)} \\
\phantom{\rho^{(3)}(f\overline g)(0)=}{}  + |1-a|^4 f'''(0) \overline{g'''(0)} \Big]  .
\end{gather*}
To~compute $[\nu]_\bm$, we~note as in Example~\ref{ex2} that for the function
$F\in\cC^\infty(\D\times\overline\D)$ given~by
\[ F(z,\ow) = z^\alpha \oz^\beta w^\gamma \ow^\delta, \qquad
 \text{where}\quad \alpha+\gamma=\beta+\delta=\rho,   \]
one has by~(\ref{e1.7A})
\[ \ps_\nu F(0) = a^\gamma \overline a^\delta   \frac{\rho!}{[\nu]_\rho} .  \]
On~the other hand, since now $\det\Phi'(0,\Lambda y)=|1-a|^2(1-|a|^2|y|^4)$
by~Lemma~\ref{lemmPhi}, we~have from~(\ref{e1.4})
\begin{gather*}
\ps_\nu F(0) =
a^\gamma \overline a^\delta |1-a|^2 \int_\D z^\rho \oz{}^\rho   \big(1-|a|^2|z|^4\big)
   \frac {(1-|z|^2)^\nu (1-|az|^2)^\nu} {|1-\overline a|z|^2|^{2\nu}}
 \frac {dz} {(1-|z|^2)^2 (1-|az|^2)^2} .
\end{gather*}
Passing to polar coordinates, we~thus obtain (writing $m$ instead of~$\rho$),
\be
 \frac{m!}{[\nu]_\bm} = |1-a|^2 \int_0^1 t^m
  \frac{(1-|a|^2 t^2)} {(1-t)^2 (1-|a|^2 t)^2}
 \frac{ (1-t)^\nu (1-|a|^2 t)^\nu} {|1-at|^{2\nu}} \, dt.   \label{e:TY}
\ee
Using series expansions, the integral can again be expressed in terms of
Horn-type two-variable hypergeometric functions, and simplif\/ies for some
special values of~$a$. In~particular, for $a=0$ the right-hand side of
(\ref{e:TY}) is~just
\[ \int_0^1 t^m (1-t)^{\nu-2} \, dt
 = \frac {m!\Gamma(\nu+1)} {\Gamma(\nu+m)},  \]
so~that
\[ \frac1{[\nu]_\bm} = \frac{\Gamma(\nu-1)} {\Gamma(\nu+m)},   \]
or,~upon renormalizing so that $[\nu]_0=1$,
\[ [\nu]_\bm = \frac{\Gamma(\nu+m)}{\Gamma(\nu)} = (\nu)_m ,  \]
in~agreement with the result
\[ \rho_\nu (f\overline g) = \sum_\bm \frac{A_\bm(f,g)}{(\nu)_\bm}   \]
from \cite{E} reviewed in Section~\ref{SecDOUBLE}. Similarly, for $a=-1$,
(\ref{e:TY})~becomes
\[ \frac1{[\nu]_\bm} = \frac{4\Gamma(2\nu-2)}{\Gamma(m+2\nu-1)} \,
 {}_2\!F_1(2\nu-1,m+1;m+2\nu-1;-1),   \]
or,~upon renormalizing $dz$ so that $[\nu]_0=1$,
\[ \frac1{[\nu]_\bm} = \frac2{(2\nu-1)_m}
 \, {}_2\!F_1(2\nu-1,m+1;m+2\nu-1;-1).   \]
Crude estimates also show that
\[ [\nu]_m \sim |1-a|^{2m} \nu^m \left[ 1 - \frac{2(1+|a|^2)}{|1-a|^2\nu}
 +O\left(\frac1{\nu^2}\right) \right]   \]
as~$\nu\to+\infty$, which can be used to check at least for the f\/irst few
terms that, again,
\[ \rho_\nu = \sum_\bm \frac{\rho^\bm}{[\nu]_\bm}   \]
is~indeed independent of~$a$, although both $\rho^\bm$ and $[\nu]_\bm$ are~not.
Note that the retraction $\pi$ in this case ($a=-1$) is simply
\[ \pi(z,\ow) = m_{z,w} ,  \]
the geodesic mid-point between $z$ and~$w$.
\end{example}

\appendix

\section{Table of parameters of real bounded symmetric domains}

The~table on the next page lists the groups $G_\R$, $K_\R$, the root
type~$\Sigma$, the~rank~$r_\R$, characteristic multiplicities $a_\R$, $b_\R$, $c_\R$
and the dimension $d$ of real bounded symmetric domains~$B_\R$, as~well as the
analogous parameters $r_\C$, $a_\C$, $b_\C$ of the complex domains~$B_\C$ and the
labellings of~$B_\C$ and~$B_\R$ following the notation
in~\cite[Chapter~11]{Loo}. The~table was mostly compiled using \cite{Zhb,Helg,FKc} and~\cite{Loo}. The~low-dimensional isomorphisms
between the various types, and the resulting restrictions on subscripts needed
in order to make the table entries non-redundant, can be found e.g.~in the
cited chapter in Loos~\cite{Loo}. As a matter of notation, we use $G_n (\Ka)$
and $U_{p,q} (\Ka)$ for the identity component of the general linear
(resp.~pseudo-unitary) group over $\Ka = \R, \C, \qH$ (=~quaternions).
$Sp_{2r}(\Ka)$ is the $2r\times 2r$-symplectic group over $\Ka = \R, \C$,
whereas $O_n(\qH)$ is the quaternion analogue of $O_n (\C)$
(usually denoted by~$SO^*(2n)$).

The~genus of $B_\C$ is given in terms of the domain parameters~by
\[ p = (r_\C-1) a_\C + b_\C + 2,   \]
while the dimension $d=\dim_\R B_\R=\dim_\C B_\C$ equals
\[ d = \frac{r_\R(r_\R-1)}2 a_\R + r_\R = \frac{r_\C(r_\C-1)}2 a_\C+r_\C  \]
for type~A, and
\[ d = r_\R(r_\R-1) a_\R + r_\R b_\R + r_\R c_\R + r_\R
= \frac{r_\C(r_\C-1)}2 a_\C + r_\C b_\C + r_\C  \]
for all other types.  Domains of type $D_2$ turn out to have,
in~some sense, two multiplicities $a$ instead of~one.

Note that the~unit interval corresponds to $I^\R_{1,1}$,
the~unit ball of~$\R^m$, $m>1$, to $I^\R_{1,m}$,
the~unit ball of $\C^m$ to $I_{1,m}$, the~unit
ball of the algebra of quaternions $\qH$  to $I^{\qH}_{2,2m}$,
the unit ball of~$\qH^m$, $m> 1$, to $I^{\qH}_{2,2m}$,
and~the unit ball of the Cayley plane $\bO^2$ to $V^\bO$.
In~the ``complex'' cases, the~root type does not quite make sense
(``BC$\times$BC'') and nor do the parameters $a_\C$, $b_\C$, $r_\C$,
while $B_\C$ is just the product $B\times\overline B$;
so~these columns are left empty.

\begin{landscape}

\vspace*{15mm}

\begin{tabular}{llcccccccccl}
$ B_\R $\quad&$ G_\R/K_\R $\quad&$ \Sigma $&$ r_\R $&$ a_\R $&$ b_\R $&$
 c_\R $&$ d $&$ r_\C $&$ a_\C $&$ b_\C $&\quad$ B_\C $\\
\noalign{\vspace{3pt}}
\hline
\noalign{\vspace{3pt}}
$ I^\R_{r,r+b} $\quad&$
 U_{r,r+b}(\R)/U_r(\R)\times U_{r+b}(\R) $\quad&$
 D_r/B_r $&$ r $&$ 1 $&$ b $&$ 0 $&$ r(r+b) $&$ r $&$ 2 $&$ b $&\quad$
 I_{r,r+b} $\\
 \noalign{\vspace{3pt}}
 $I_{r,r+b}$ &$ U_{r,r+b}(\C)/U_r(\C)\times U_{r+b}(\C) $\quad&$
  $&$ r $&$ 2 $&$ 2b $&$ 1 $&$ 2r(r+b) $&$  $&$  $&$  $&\quad (product case)\\
\noalign{\vspace{3pt}}
$ I^\qH_{2r,2r+2b} $\quad&$
 U_{r,r+b}(\qH)/U_r(\qH)\times U_{r+b}(\qH) $\quad&$
 C_r/BC_r $&$ r $&$ 4 $&$ 4b $&$ 3 $&$ 4r(r+b) $&$ 2r $&$ 2 $&$ 2b $&\quad$
 I_{2r,2r+2b}$\\
  \noalign{\vspace{3pt}}
$ V^{\bO_0} $\quad&$
 U_{2,2} (\qH)/U_{2} (\qH)\times U_{2} (\qH) $\quad&$
 B_2 $&$ 2 $&$ 3 $&$ 4 $&$ 0 $&$ 16 $&$ 2 $&$ 6 $&$ 4 $&\quad$
   V$ \\
 \noalign{\vspace{3pt}}
$ III^\R_r $\quad&$
 G_r (\R)/U_r (\R) $\quad&$
 A_r $&$ r $&$ 1 $&$ - $&$ - $&$ \frac12 r(r+1) $&$ r $&$ 1 $&$ 0 $&\quad$
  III_r$\\
 \noalign{\vspace{3pt}}
$ I^\C_{r,r} $\quad&$
 G_r (\C)/U_r (\C)$\quad&$
 A_r $&$ r $&$ 2 $&$ - $&$ - $&$ r^2 $&$ r $&$ 2 $&$ 0 $&\quad$
  I_{r,r}$\\
 \noalign{\vspace{3pt}}
$ II^\qH_{2r} $\quad&$
 G_r (\qH)/U_r (\qH) $\quad&$
 A_r $&$ r $&$ 4 $&$ - $&$ - $&$ r(2r-1) $&$ r $&$ 4 $&$ 0 $&\quad$
  II_{2r}$\\
  \noalign{\vspace{3pt}}
$ VI^{\bO_0} $\quad&$
 G_4 (\qH)/U_4 (\qH) $\quad&$
 D_3 $&$ 3 $&$ 4 $&$ 0 $&$ 0 $&$ 27 $&$ 3 $&$ 8 $&$ 0 $&\quad$
  VI$\\
 \noalign{\vspace{3pt}}
$III_r$  &$ Sp_{2r}  (\R)/U_r (\C) $\quad&$
  $&$ r $&$ 1 $&$ 0 $&$ 1 $&$ r(r+1) $&$  $&$  $&$  $&\quad (product case)\\
  \noalign{\vspace{3pt}}
$ III^\qH_{2r} $\quad&$
 Sp_{2r} (\C)/U_r (\qH) $\quad&$
 C_r $&$ r $&$ 2 $&$ 0 $&$ 2 $&$ r(2r+1) $&$ 2r $&$ 1 $&$ 0 $&\quad$
  III_{2r}$\\
\noalign{\vspace{3pt}}
$ II^\R_{2r+\varepsilon} $\quad&$
 O_{2r+\varepsilon} (\C)/U_{2r+\varepsilon} (\R) $\quad&$
 D_r/B_r $&$ r $&$ 2 $&$ 2\varepsilon $&$ 0 $&$ r(2(r+\varepsilon)-1) $&$ r $&$ 4 $&$ 2 $&\quad$
  II_{2r+\varepsilon}$\\
 \noalign{\vspace{3pt}}
$II_{2r+\varepsilon}$ &$  O_{2r+\varepsilon} (\qH)/U_{2r+\varepsilon} (\C) $\quad&$
  $&$ r $&$ 4 $&$ 4\varepsilon $&$ 1 $&$ 2r(2(r+\varepsilon)-1) $&$  $&$  $&$  $&\quad (product case)\\
\noalign{\vspace{3pt}}
$  IV^{\R,q}_{p+q} $\quad&$
 SO_{p,1}\times SO_{1,q}/SO_{p,0}\times SO_{0,q} $\quad&$
 D_2/A_2 $&$ 2 $& n/a &$ 0 $&$ 0 $&$ p+q $&$ 2 $&$ p+q-2 $&$ 0 $&\quad$
 IV_{p+q}$\\
 \noalign{\vspace{3pt}}
 $IV_n$ &$ SO_{n,2}/SO_{n,0}\times SO_{0,2} $\quad&$
  $&$ 2 $&$ n-2 $&$ 0 $&$ 1 $&$ 2n $&$  $&$  $&$  $&\quad (product case)\\
   \noalign{\vspace{3pt}}
$V$ &$ E_{6(-14)}/Spin(10)\times SO(2) $\quad&$
  $&$ 2 $&$ 6 $&$ 8 $&$ 1 $&$ 32 $&$  $&$  $&$  $&\quad (product case)\\
\noalign{\vspace{3pt}}
$ IV^{\R,0}_n$\quad&$
 SO_{n,1}/SO_{n,0} $\quad&$
 C_1 $&$ 1 $&$ - $&$ 0 $&$ n-1 $&$ n $&$ 2 $&$ n-2 $&$ 0 $&\quad$
  IV_n$\\
    \noalign{\vspace{3pt}}
$ V^\bO $\quad&$
 F_{4(-20)}/SO(9) $\quad&$
 BC_1 $&$ 1 $&$ - $&$ 8 $&$ 7 $&$ 16 $&$ 2 $&$ 6 $&$ 4 $&\quad$
 V$\\
\noalign{\vspace{3pt}}
$VI$ &$ E_{7(-25)}/E_6\times SO(2) $\quad&$
  $&$ 3 $&$ 8 $&$ 0 $&$ 1 $&$ 54 $&$  $&$  $&$  $&\quad (product case)\\
\noalign{\vspace{3pt}}
$ VI^\bO  $\quad&$
 E_{6(-26)}\times O(2)/F_4\times O(1) $\quad&$
 A_3 $&$ 3 $&$ 8 $&$ - $&$ - $&$ 27 $&$ 3 $&$ 8 $&$ 0 $&\quad$
 VI$\\
\noalign{\vspace{3pt}}
\hline
\end{tabular}
\end{landscape}

\subsection*{Acknowledgements}
Research supported by the German-Israeli Foundation (GIF), I-696-17.6/2001;
the Academy of Sciences of the Czech Republic institutional research plan
no.~AV0Z10190503; and GA~{\v C}R grant no.~201/06/0128.

\pdfbookmark[1]{References}{ref}
\LastPageEnding

\end{document}